\documentclass[twocolumn]{aastex631}

\usepackage{amsmath,amsfonts,amssymb}
\usepackage{graphicx}
\usepackage{enumitem}
\usepackage{gensymb}
\usepackage{csquotes}
\usepackage{array}
\usepackage{booktabs}
\usepackage{microtype}

\usepackage{savesym}
 \savesymbol{tablenum}
\usepackage{siunitx}
 \restoresymbol{SIX}{tablenum}
\usepackage{xpatch}
\xpatchcmd{\thebibliography}{\twocolumngrid}{}{}{}
\journalinfo{}
\makeatletter
\let\frontmatter@title@above=\relax
\makeatother

\def\Plus{\texttt{+}}

\begin{document}

\title{Tuning the MAPS Adaptive Secondary Mirror: Actuator Control, PID Tuning, Power Spectra and Failure Diagnosis}

\author{Jess A. Johnson}
 \affiliation{Steward Observatory, University of Arizona, 933 N. Cherry Ave., Tucson AZ 85721, USA}
\author{Amali Vaz}
 \affiliation{Steward Observatory, University of Arizona, 933 N. Cherry Ave., Tucson AZ 85721, USA}
\author{Manny Montoya}
 \affiliation{Steward Observatory, University of Arizona, 933 N. Cherry Ave., Tucson AZ 85721, USA}
\author{Katie M. Morzinski}
 \affiliation{Steward Observatory, University of Arizona, 933 N. Cherry Ave., Tucson AZ 85721, USA}
\author{Jennifer Patience}
 \affiliation{School of Earth and Space Exploration, Arizona State University, Tempe, AZ 85281, USA}
\author{Suresh Sivanandam}
 \affiliation{David A. Dunlap Department of Astronomy \& Astrophysics, University of Toronto}
 \affiliation{Dunlap Institute for Astronomy and Astrophysics, University of Toronto}
\author{Guido Brusa}
 \affiliation{Large Binocular Telescope, Steward Observatory, 933 N. Cherry Ave., Tucson, AZ 85721, USA}
\author{Olivier Durney}
 \affiliation{Steward Observatory, University of Arizona, 933 N. Cherry Ave., Tucson AZ 85721, USA}
\author{Andrew Gardner}
 \affiliation{Steward Observatory, University of Arizona, 933 N. Cherry Ave., Tucson AZ 85721, USA}
\author{Olivier Guyon}
 \affiliation{Steward Observatory, University of Arizona, 933 N. Cherry Ave., Tucson AZ 85721, USA}
\author{Lori Harrison}
 \affiliation{Steward Observatory, University of Arizona, 933 N. Cherry Ave., Tucson AZ 85721, USA}
\author{Ron Jones}
 \affiliation{Steward Observatory, University of Arizona, 933 N. Cherry Ave., Tucson AZ 85721, USA}
\author{Jarron Leisenring}
 \affiliation{Steward Observatory, University of Arizona, 933 N. Cherry Ave., Tucson AZ 85721, USA} 
\author{Jared Males}
 \affiliation{Steward Observatory, University of Arizona, 933 N. Cherry Ave., Tucson AZ 85721, USA}
\author{Bianca Payan}
 \affiliation{Steward Observatory, University of Arizona, 933 N. Cherry Ave., Tucson AZ 85721, USA}
\author{Lauren Perez}
 \affiliation{Steward Observatory, University of Arizona, 933 N. Cherry Ave., Tucson AZ 85721, USA}
\author{Yoav Rotman}
 \affiliation{School of Earth and Space Exploration, Arizona State University, Tempe, AZ 85281, USA}
\author{Jacob Taylor}
 \affiliation{David A. Dunlap Department of Astronomy \& Astrophysics, University of Toronto}
 \affiliation{Dunlap Institute for Astronomy and Astrophysics, University of Toronto}
\author{Dan Vargas}
 \affiliation{Steward Observatory, University of Arizona, 933 N. Cherry Ave., Tucson AZ 85721, USA}
\author{Grant West}
 \affiliation{Steward Observatory, University of Arizona, 933 N. Cherry Ave., Tucson AZ 85721, USA}

\let\thefootnote\relax\footnotetext{Further author information: \\J.A.J.:  jajohnson@arizona.edu, 1 520 621-9519 \\ A.V.:  amali@arizona.edu, 1 520 621-2288}


\begin{abstract}
The MMT Adaptive optics exoPlanet characterization System (MAPS) is currently in its engineering phase, operating on sky at the MMT Telescope on Mt. Hopkins in southern Arizona. The MAPS Adaptive Secondary Mirror’s actuators are controlled by a closed loop modified PID control law and an open loop feed forward law, which in combination allows for faster actuator response time. An essential element of achieving the secondary’s performance goals involves the process of PID gain tuning. To start, we briefly discuss the design of the MAPS ASM and its actuators. We then describe the actuator positional control system and control law. Next, we discuss a few of the issues that make ASM tuning difficult. We then outline our initial attempts at tuning the actuator controllers, and discuss the use of actuator positional power spectra for both tuning and determining the health and failure states of individual actuators. We conclude by presenting the results of our latest round of tuning configuration trials, which have been successful at decreasing mirror latency, increasing operational mirror modes and improving image PSF. 

\end{abstract}



\section{INTRODUCTION}

\subsection{History of the MAPS project}

The MMT Adaptive Optics Exoplanet Characterization System, or MAPS, is the evolutionary descendent of a long line of experimental prototypes and telescope adaptive optics systems. In 1998, the six mirrors of the Multi Mirror Telescope (MMT) were replaced by a single 6.5m monolithic mirror, and a unique adaptive optics system was designed to accompany the upgrade. The MMT Adaptive Optics system (MMTAO) was a first-of-its-kind design, as it introduced a revolutionary idea: wavefront correction by a large deformable mirror positioned at a telescope's secondary mirror position.  The MMT's \emph{adaptive secondary mirror}, or ASM, was a first generation device that introduced an entirely new system architecture for large telescope adaptive optics \citep{Morzinski20}.

The development program for an AO system utilizing adaptive secondaries began in 1992 and was a cooperative effort between private companies and research institutions in both Italy (Arceti Observatory) and the United States (University of Arizona). Most of the theory and electronics design was done by Italian companies, of which perhaps Microgate is the most well-known, and the technology for developing thin-shell mirrors was done at the University of Arizona's Steward Observatory. 

The MMTAO, with its ground-breaking 336 voice coil actuator ASM (MMT336), was decommissioned in 2017. Its replacement, what was to become MAPS, was initially funded by an NSF Mid-Scale Innovations Program in Astronomical Sciences (MSIP) seed grant issued in August of 2018. Unlike the original MMT system, MAPS was to be both designed and built in-house at the University of Arizona's Steward Observatory. 

Development and fabrication proceeded on schedule, until the pandemic forced the closure of the University. Multiple departures and retirements among key personnel involved in the design of the ASM and its control systems resulted in the loss of first hand insight and experience, and interrupted the creation of operational documentation.  This resulted in the MAPS ASM going on-sky with its \emph{actuator control system} (ACS) in its default state and with little knowledge about its operation or optimization. And because an adaptive secondary is a rather uncommon device (a magnetically levitated thin glass mirror controlled by hundreds of autonomous positional control actuators), standard industrial control theory and its tuning methods, which don't take into account considerations such as mirror modes and oscillatory fragility, have only limited applicability. 

The MAPS ASM tuning project starts with the ACS being, basically, a black box.

\subsection{Design of the MAPS ASM}

Most of the refinements and improvements in AO performance that distinguish MMTAO from MAPS relate directly to substantial changes in the design of the ASM. This paper is concerned with the adaptive secondary's actuators that control the surface figure of the mirror, and the positional systems that control them (ACS), and this is one of the many areas of substantial change. A brief overview of the structure of the ASM and the design of its actuators will be helpful in understanding what follows. For a complete discussion of the mechanical and electrical design of the ASM, refer to \citet{johnson23}.

\subsubsection{Design criteria}

MAPS was conceived of as an exoplanet characterization system, and science goals largely drove the design of the ASM. In comparison to the MMT336, MAPS requires higher spectral resolution, an increase in achievable contrast, improved image quality over a broader wavelength range, and increased throughput and operational efficiency. The following were the performance requirements that informed the design, given as comparisons to MMT336 \citep{Hinz18,Vaz20}:

\begin{itemize}[noitemsep]
  \item An operating bandwidth of 1000 Hz.
  \item Positional calculations at 1kHz.
  \item Temporal lag reduction and a decrease in mirror settling time by a factor of 2-3.
  \item The implementation of an open loop feed-forward component to actuator control.
  \item A decrease in the system update rate from 2.7 ms lag to 1 ms.
  \item 120 corrected modes, up from 55.  
\end{itemize}

With these changes, it was estimated that residual wavefront error could be reduced from between 400 and 500 nms RMS for the old system to 200 nm RMS for the new system, with corresponding gains in Strehl ratio. The ability to achieve these goals was only partially achieved by electronics and hardware changes; even more important were the changes to the actuator control law. 

\subsubsection{Description of the MAPS ASM}
\label{sec:description}

\begin{figure}[!t]
    \centering
        \includegraphics[height=2.5in]{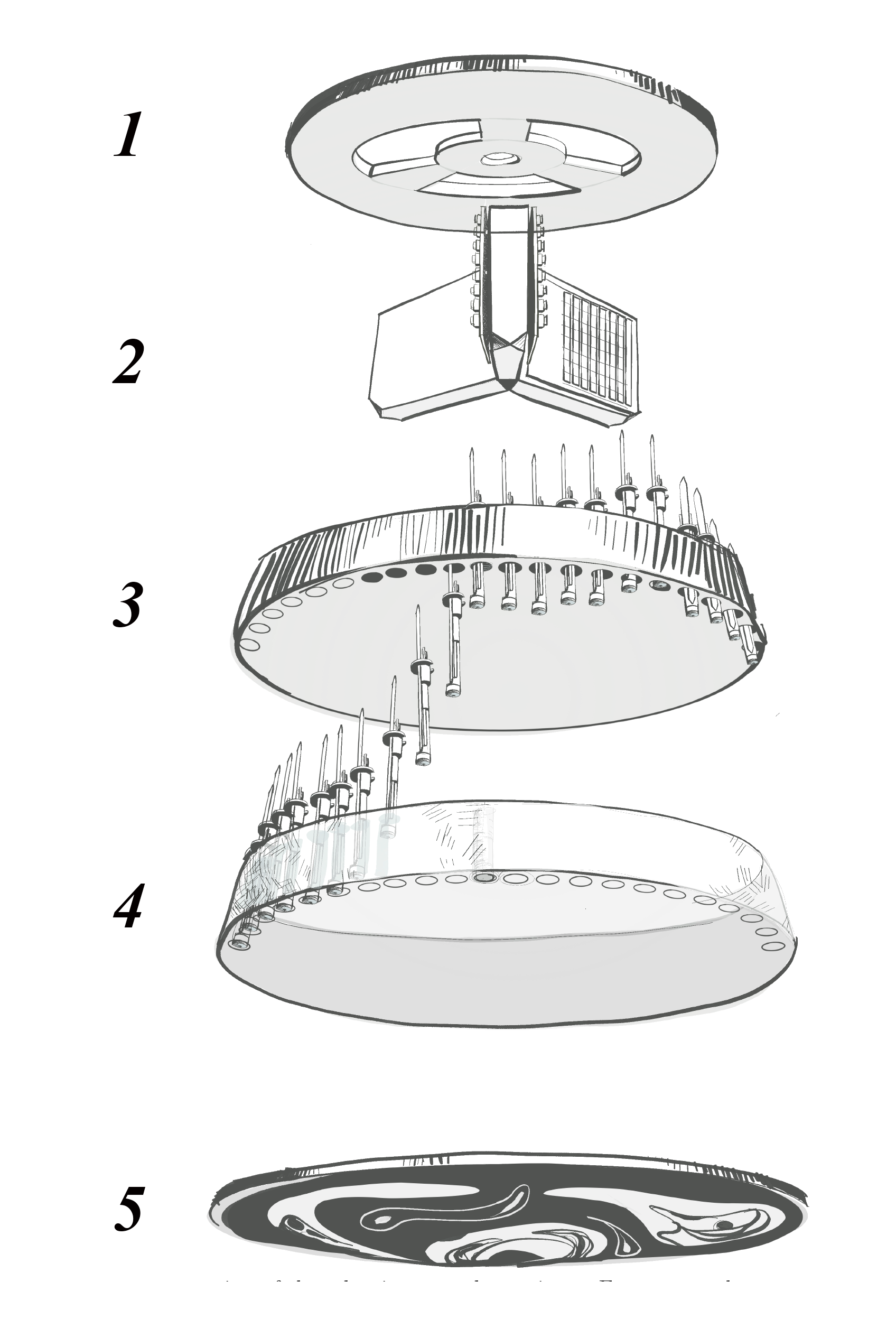}
    \caption{A blowup diagram of the MAPS adaptive secondary mirror, showing its major structural components. 1) Ring interface, 2) Central electronics hub, 3) Cold plate, 4) Reference body, and 5) Thin shell. \protect\citep{Vaz20}}
    \label{fig:ASMBlowup}
\end{figure}

The MMT telescope is a Cassegrain design with a 6.5-m f/1.25 primary mirror. The adaptive secondary is an f/15 deformable convex mirror, which provides a 20 arcminute field of view \citep{West97}. A blowup diagram of the ASM is shown in Figure~\ref{fig:ASMBlowup}.

The five primary components of the ASM as indicated in the diagram are as follows: 

The \textbf{reference body} is  composed of a monolithic \SI{500}{\milli\metre} thick piece of Zerodur glass, through which 336 boreholes have been drilled to house actuators. It provides a stable reference surface for determining mirror position. The \textbf{cold plate} is a single piece of aluminum which serves as a surface to which the actuators attach and provides a passive thermal sink. The \textbf{hexapod and telescope interface} is a mounting system attached to the reference body through the cold plate; its other end is fixed to a metal electronics frame and ring interface which mounts to the telescope. 

The \textbf{electronics hub} holds the system's housekeeping board, control motherboard, and six daughterboards; each daughterboard is attached to 56 actuators via USB-C cables. Commands, actuator status info, and monitoring flow between this architecture and a control computer via Ethernet. And the \textbf{thin shell} is the secondary mirror's reflective surface. It is 64 cm in diameter, made of Zerodur glass that ranges from 1.8 to \SI{1.9}{\milli\metre} in thickness and weighs \SI{2.6}{\kilo\gram}. Attached to the upper surface of the thin shell are small, radially polarized rare earth magnets positioned directly above each of the boreholes. The magnetic field generated by the actuator coil interacts with these magnets to provide the force that moves the mirror.
 
\subsubsection{MAPS actuators}
\label{sec:actuators}

The MAPS actuators represent a considerable upgrade in both electronics and functionality over their predecessors, but the basic procedure they accomplish is the same: the distance between the reference body and the portion of the mirror over an actuator (referred to in this discussion as a \emph{mirror segment}) embedded in a reference body borehole is measured by the capacitive sensing system, described in Sec.~\ref{sec:capsensing} below. The difference between the mirror segment's measured position and its desired position (provided by the AO system's reconstructor in response to measurements taken by the wavefront sensor) is then calculated, and the result is fed into the actuator's onboard ACS controller. The controller's output is converted to a current value, and the current is sent through the actuator's coil, creating the magnetic force which in turn moves the mirror.

The MAPS actuator design consolidates almost every fundamental ASM function to its onboard electronics \citep{Downey19}, unlike its predecessor which relied on external electronics to do positional and current calculations. The functions performed at actuator level are:

\begin{itemize}[noitemsep]
  \item Measurement of the capcitive decay voltage;
  \item Digitizing of the measurement;
  \item Calculation of the force required to move the mirror;
  \item Implementation of the modified PID control law;
  \item Production of the coil current required to produce the desired magnetic field;
  \item The execution of safety and status checks on the actuator..  
\end{itemize}

This concentration of functionality leads to a decentralized system of actuator control, in which all actuators act independently of each other, as opposed to being commanded by a centralized control computer (see Section~\ref{sec:structure}).

\subsubsection{Capacitive sensing system}
\label{sec:capsensing}

\begin{figure}[!t]
    \centering
        \includegraphics[height=2.18in]{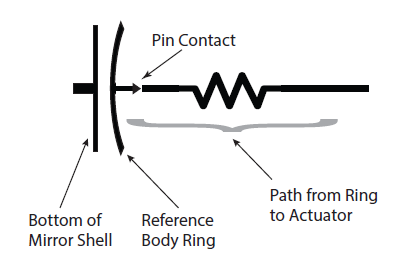}
    \caption{Schematic of the MAPS actuator capacitive sensing system. A voltage pulse is sent through the conductive bottom of the thin shell mirror, which produces a voltage at the reference body ring surrounding each actuator. The RC circuit produces a decay curve which is read by the actuator electronics and converted to a distance measurement,  }
    \label{fig:capsensor}
\end{figure}

A core component of the MAPS actuators is the capacitive sensing system. This system extracts a distance measurement by using the geometry of a parallel plate capacitor, where the bottom of the mirror acts as one plate and a metallic circular ring on the reference body encircling the actuator bore hole acts as the other (see Figure~\ref{fig:capsensor}). As the mirror moves, the distance between the plates changes, altering the capacitance. The voltage between the plates is directly proportional to this distance and indirectly proportional to the capacitance. 

To measure the displacement of the mirror from the reference body, a voltage pulse, called the `Go' pulse, is sent across the conductive surface on the underside of the thin shell. This pulse flows across each actuator's associated thin-shell/reference-body capacitor and is then measured by the actuator. The voltage drops off as an RC decay curve, and by reading this voltage at two set times on the curve, the actuator can determine the equivalent displacement of the mirror. The system uses a 16-bit ADC which provides \SI{3}{\nano\metre} resolution over a \SI{197}{\micro\metre} throw length.

\subsubsection{Fundamental quantities}
\label{sec:qunatities}

This brief review of the ASM and its actuators gives us a overview of the quantities that we will be concerned about in this discussion. It also allows us to re-frame these components in the language of \emph{control theory}:

The thin shell mirror is the \emph{controlled object}, or \emph{plant}. The distance between the reference body and the thin shell, measured by the capacitive sensing system (the \emph{sensor}), is the quantity being controlled, called the \emph{process variable}. The position we want the portion of the mirror over any individual actuator to be located at is called the \emph{set point}. The difference between the desired set point and the momentary value of the process variable is the \emph{error}. The calculated value of the current that creates the magnetic force applied by the actuator coil that moves the mirror and corrects the error is the \emph{process output}. The device that applies the quantity that causes the change to the plant is called an \emph{actuator}. And finally, the electronics that control this whole process is called the \emph{controller}.


\section{ADAPTIVE SECONDARY MIRRORS \& THE ACTUATOR CONTROL PROBLEM}
\label{sec:ASMControl}

Before discussing the MAPS controller and the process of tuning the MAPS ASM, it is important to understand how mirror control is accomplished, and what constraints and considerations affect the process. 

Adaptive secondary mirror systems cleverly solve one adaptive optics design issue, but in the process of doing so, create another, more difficult issue. Conventional AO systems, in the process of rerouting a portion of the telescope’s beam to a small deformable mirror located on a breadboard-based system, introduce multiple optical surfaces into the light path, adding unwanted polarization, emissivity, and reflective loss to the light that is used to determine the AO system’s correction. Adaptive secondaries solve this issue by placing the deformable mirror directly in the telescope beam.

However, these large and complicated mirrors are controlled by hundreds of actuators with more complex electronics and functionality than their smaller counterparts in traditional deformable mirrors. Limitations in both computational power and communication speed have, to this point, mandated control systems designs in which actuators behave largely independently of each other. A large number of autonomously acting actuators independently exerting force on the same fragile piece of glass creates what has come to be known in the literature as the \emph{actuator control problem}. 

\subsection{The actuator control problem}
\label{sec:ACP}

This problem has two components. The first is: how to design an actuator control system capable of applying enough constrained force to a small region of fragile glass so that it can rapidly assume its desired position \emph{without} substantially overshooting or threatening mirror integrity? 

This alone is a complex problem, given that the desired mirror position is constantly changing in response to the input from the wavefront sensor, and that actuator behavior, while driven by an autonomous process, is not truly independent but coupled through the mirror function.

The second component concerns an issue unique to large magnetically levitated thin shell mirrors, and that is that they have oscillatory states, called Eigenmodes, which are similar to standing waves, resulting from there being no mechanical contact to the mirror to provide damping. Conventional deformable mirrors change their surface figures only in response to force applied by actuators (the \emph{quasi-static modeling assumption}). But levitated thin shell mirrors are free to change their surface figures in response to more than just actuator force (the \emph{dynamic modeling assumption}). This complicates the control situation: `\emph{...[another] problem to solve is that posed by a system, the deformable mirror, that has a very high density of resonant frequencies over a very wide frequency range'} \citep{salinari98}. For more on this, see Section~\ref{sec:resonances}.

\subsection{Dynamics and positional control}

To address the first part of the control problem, we consider the dynamics of the situation. The problem, at first glance, seems deceptively simple. We have an object that has to be moved to a predetermined position, either forward or backward from its current position, and a device that issues a force that can either push or pull on the object. The question becomes, what is the most appropriate way to apply the force?

To move the mirror segment, we look at the displacement required to reach the desired position and calculate a current that, when applied to the mirror's magnet, accelerates the segment of the mirror in the desired direction. In theory we could determine exactly the amount of force required to accelerate the mirror to travel the appropriate distance, allow the restoring force supplied by the mirror to decelerate the motion, and let the segment come to a stop exactly where desired. We could then calculate exactly the amount of current required to produce this force. In real life, however, this is an incredibly complicated calculation due to the number and nature of the forces acting on the mirror. The mirror resists the motion with a type of spring force; the mirror may be in motion already from either previous motion or Eigenmode oscillation; air molecules resist the free motion of the mirror; and so on.

\begin{figure*}[!t]
    \centering
        \includegraphics[height=3.2in]{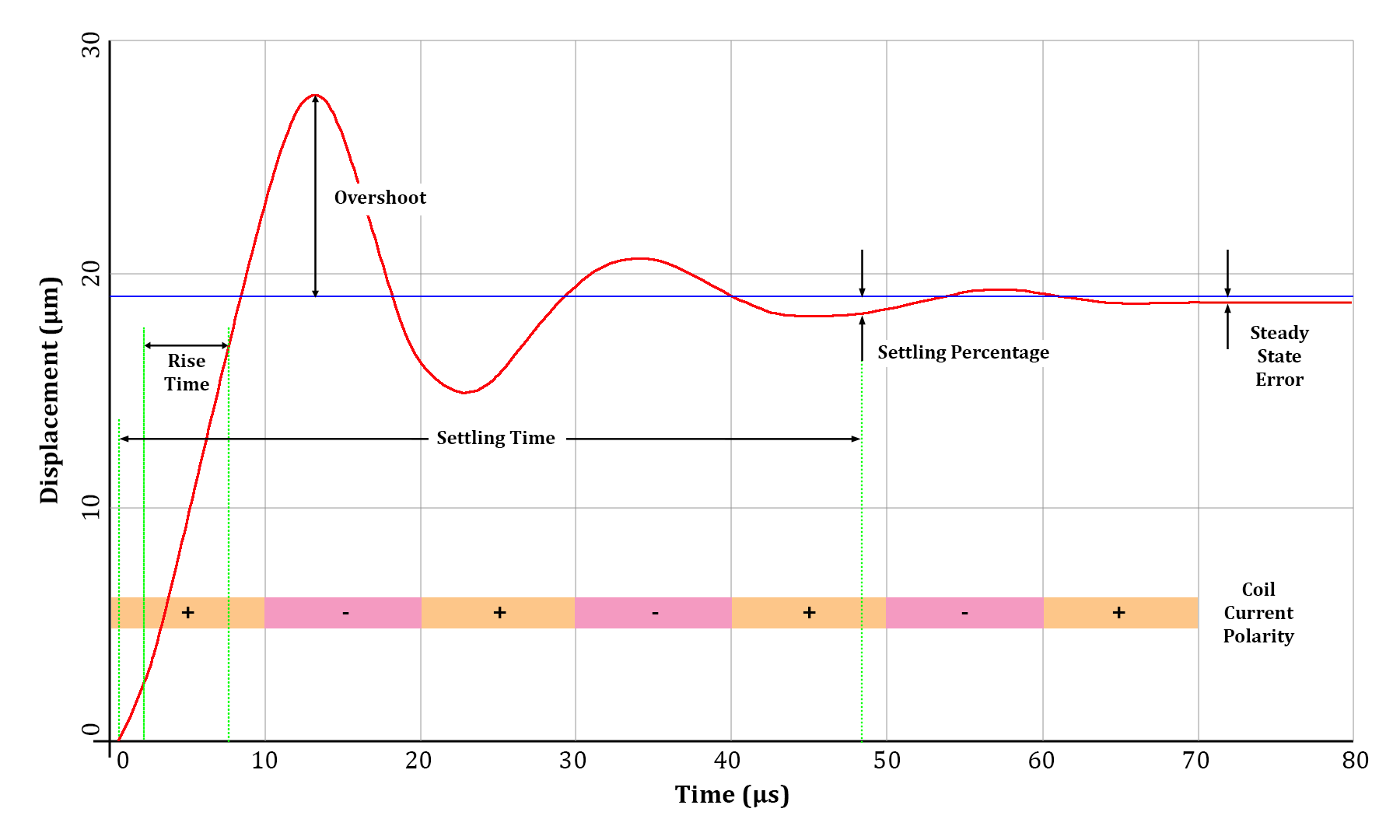}
    \caption{Hypothetical plot of Proportional-Integral-Derivative action on a mirror segment. The red line indicates the mirror's displacement vs time. The blue line is the mirror's desired position. The bar across the bottom of the graph indicates the polarity of the current being applied to the actuator coil. The arrows and associated labels indicate fundamental PID quantities.}
    \label{fig:oscillation}
    \vspace{4mm}
\end{figure*} 

This means that we have to give up on the idea of moving a mirror segment directly to its desired location, and instead create a form of damped oscillatory motion in which we apply a force that is greater than needed, sending the mirror past its desired position, then reverse the applied force, pulling the mirror in the opposite direction, then reverse the force again, and so on. Each time we do this, the mirror gets closer to its desired position, oscillating around it until it eventually stops, within a small margin of error, at its targeted position. This is illustrated in Figure~\ref{fig:oscillation}. The device that controls this process is called a \emph{Proportional-Integral-Derivative controller}, or PID.

\vspace{1.1mm}
The figure shows a displacement vs time plot of a mirror segment's physical location. At $t=0$ the mirror segment is at its initial location. The blue line indicates the displacement from its initial position required to reach its desired location, or \emph{set point}. The bar across the bottom of the plot indicates the polarity of the current being applied to the coil, with `+' denoting the initial polarity creating motion towards the set point.

\vspace{1.24mm}
At the start of the positional cycle, current is applied to the coil, creating a magnetic force that accelerates the mirror segment towards its set point. The amount of current that is applied is calculated by the PID controller. \SI{10}{\micro\second} later, as the segment overshoots the set point, the next PID cycle begins with the set point in the opposite direction, and the controller again calculates the current required, and the segment decelerates, stops, reverses direction, and again overshoots the set point. 
\vspace{1.2mm}

The \emph{rise time} is the amount of time it takes for the segment to move from $10\%$ to $90\%$ of the set point. The \emph{overshoot} is the distance that the segment moves past the set point during its initial motion. This oscillatory behavior repeats until the segment eventually comes within an acceptable error margin from the set point called the \emph{steady state error}. The \emph{settling time} is the amount of time it takes to reach a defined error margin, which for MAPS is set at 10\% of the desired position. Decreasing the settling time is the goal of optimization.

This process happens simultaneously but independently for all 336 actuators on the mirror, and the current/force calculations are determined by the modified PID controller built into each actuator's electronics. The PID controller bases its calculations on the start-of-cycle segment position, the initial reconstructor command, and the values of control variables that have been determined and set during the process called \emph{tuning}.

\subsection{Result of tuning: settling time and mirror modes}
\label{sec:settling}

Each cycle of mirror segment positioning requires the controller to move the mirror segment utilizing magnetic force, and the amount of time it takes for the mirror segment to reach its set point is that actuator's \emph{settling time}. Each actuator's settling time is largely determined by its tuning. In the bigger picture, every actuator must position its corresponding mirror segment so that the overall surface figure of the mirror corresponds to the reconstructor's correction. The time it takes to accomplish this is the \emph{mirror settling time}. 

For the MAPS ASM, the reconstructor, acting on input from the wavefront sensor, sends a complete set of 336 actuator position commands every \SI{1}{\milli\second}; this is the controller's \emph{outer loop timing}. The PID controller operates at 100 times this speed; one PID cycle takes \SI{10}{\micro\second} to complete. This is the controller's \emph{inner loop timing}. If any particular actuator fails to settle its associated mirror segment before the next outer loop cycle begins, the ability of the entire mirror to effectively correct the wavefront is degraded; the more mirror segments that fail to settle, the worse the wavefront correction becomes. 

This also has ramifications on the number of \emph{mirror modes} the mirror can produce. Mirror modes are similar to basis sets such as the Zernike polynomials, but are unique to each ASM; they are the fundamental components of wavefront correction. Lower order aberrations, such as tip/tilt and focus, which rely on modes composed of large groups of actuators acting similarly, can still be partially corrected if some of those actuators miss their targets. Higher order aberrations, corrected by modes composed of higher spatial frequencies, require the independent positioning of actuators at widely varying locations and displacements to effectively apply a correction. If the actuators involved miss their set points, the mirror becomes ineffectual at higher order correction. MAPS running under default PID tuning produced a maximum of 50 modes. MAPS with modified tuning was able to correct up to 100 modes. We expect that MAPS will reach 120 modes safely under optimal tuning. 

\subsection{Oscillation and mirror resonances}

Tuning actuators that push and pull on a thin sheet of glass is an inherently fraught proposition, which goes to the first aspect of the actuator control dilemma (see the introduction to  Section~\ref{sec:ACP}). Put simply, if we apply force to the mirror inappropriately, we can shatter it. 

This can happen in two ways. The first is by miscalculating the amount of overshoot needed at the first stage of the positional control cycle, i.e., we apply too much force, overcoming the material properties of the glass and cracking it. Since we basically determine the amount of overshoot during tuning, this is our first serious constraint. The second, and more subtle way we can threaten mirror integrity is through uncontrolled oscillation. 

\subsubsection{PID oscillation}
\label{subsubsec:WildCoil}

The PID process is inherently oscillatory, and as long as the control variables are set correctly during tuning, the overshoot amplitude can be controlled and the oscillation dampens quickly. This behavior is not threatening and is indeed the way PID controllers are designed.

But a PID can accidentally be tuned to cause uncontrolled oscillation, specifically by setting either the proportional gain (PG) too high, or by incorrect adjustment of the derivative gain (DG). This is called \emph{unstable tuning}. This can become an issue if the operational range of these variables is not known in advance, as was the case with the MAPS controller. Because the values of these variables can range over seven orders of magnitude, even picking an incorrect starting value to experiment with can hinder discovering the correct values at best, or at worst cause the PID to go unstable as soon as the first cycle starts.

Adding to this issue is an inherent limitation in the design of the MAPS actuators. Certain tuning states can cause the actuator to exhibit a behavior called WildCoil, in which the H-bridge (an electronic component designed to switch the direction of current flow to the coil) begins to oscillate current polarities, forcing the magnetic field to alternate rapidly back and forth. If the initial magnitude of the calculated current is too high, the mirror can shatter as a result. To counter this, the housekeeping circuitry of the actuator is designed to send a distress signal to the main control computer, `safing' the mirror, the equivalent of system shutdown. But even here, the threshold for WildCoil detection is another variable to be configured. Set the WildCoil value too low, and the system shuts down in response to nominal H-bridge behavior. Set the value too high, and the actuator becomes blind to this potentially catastrophic behavior.

\begin{table}[b!]
    \vspace{3mm}
    \centering
    \caption{Resonance Modes of the MAPS Mirror, from \citet{grocott97}}
      \begin{tabular}{c r c l c c} 
        \toprule
        Mode & Freq & Node & Type & Num & Group \\
        (\#) & (Hz) & ($\theta,r$) & & & (j) \\
        \midrule
        01 & 7.0 & (1,1) & Circumferential & 2 & 1 \\ 
        02 & 31.3 & (2,0) & Circumferential & 2 & 2  \\
        03 & 32.2 & (0,0) & Radial (Piston) & 1 & 0  \\
        04 & 72.5 & (3,0) & Circumferential & 2 & 3  \\
        05 & 89.7 & (0,0) & Torsion  & 1 & 0  \\
        06 & 126.6 & (4,0) & Circumferential & 2 & 2  \\
        07 & 192.3 & (5,0) & Circumferential & 2 & 1  \\
        08 & 269.1 & (6,0) & Circumferential & 2 & 0  \\
        09 & 358.0 & (7,0) & Circumferential & 2 & 1  \\
        10 & 458.3 & (8,0) & Circumferential & 2 & 2  \\
        11 & 460.2 & (0,1) & Radial & 1 & 0  \\
        12 & 464.8 & (1,1) & Circum-Radial & 2 & 1  \\
        13 & 481.7 & (2,1) & Circum-Radial & 2 & 2  \\ 
        14 & 520.4 &  (0,2) & Radial & 1 & 0  \\
        15 & 533.9 & (3,1) & Circum-Radial & 2 & 3  \\
       \bottomrule
     \end{tabular}
   \label{table:modes}
\end{table} 

\subsubsection{Mirror Resonances}
\label{sec:resonances}

The thin shell mirror is essentially levitated in a magnetic field. Because of this, there is no physical damping and the mirror is free to vibrate. It does this in response to any force that is applied to it, from the action of the magnetic coil actuators, to the motion of air in the gap between it and the reference body, and to vibration emanating from the motion of the telescope or the movement of the dome. Dynamic modeling shows that this vibration produces a variety of standing wave patterns, known as \emph{mirror resonances} or \emph{eigenmodes}. 

\begin{figure*}[t!]
    \vspace{2mm}
    \centering
        \includegraphics[height=1.15in]{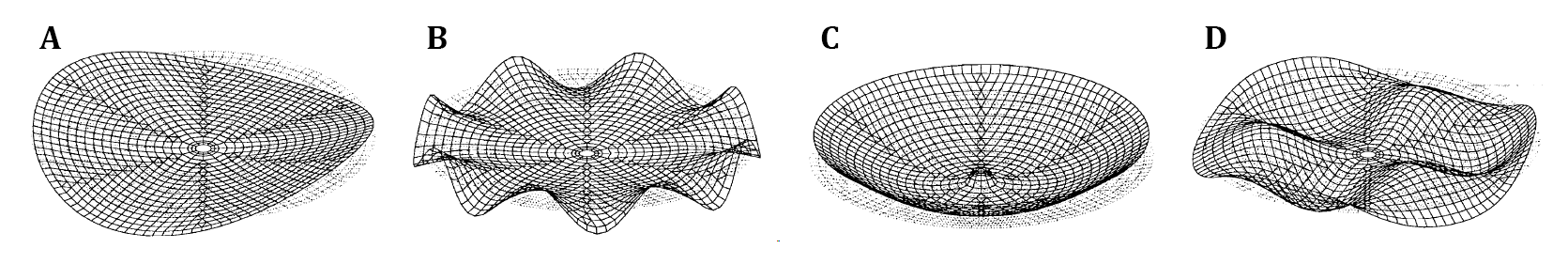}
    \caption{Mirror resonance modes of the MAPS thin-shell mirror. A) Mode 2: 31.3Hz $[2-\theta]$; B) Mode 10: 458.3Hz $[8-\theta]$; C) Mode 11: 460.2Hz $[1-r]$; D) Mode 15: 533.9 Hz $[3-\theta,1-r]$. Graphics from \protect\citet{grocott97}.}
    \label{fig:modes}
    \vspace{4mm}
\end{figure*}

Finite element analysis of the MAPS mirror identifies thirty natural resonance modes ranging from \SI{7}{\hertz} to \SI{1}{\kilo\hertz}, of which the first fifteen are listed in Table~\ref{table:modes}, along with the frequency at which they occur. Each of these modes is a unique pattern of vibration, a few of which are illustrated in Figure~\ref{fig:modes}.

\subsubsection{Rigid vs non-rigid actuators}

Voice coil actuators, operating via applied magnetic force to a magnet affixed to the rear of a portion of a thin shell deformable mirror, are by nature \emph{non-rigid}. In this usage, non-rigid means that the actuator itself is not physically affixed to the mirror. Rigid actuators act as a damping force that attenuates or eliminates resonant modes below a threshold frequency. 

Without the damping of rigid actuators, a voice coil actuated system requires another source of strong damping, and this is primarily provided by the viscous damping effect of air in the gap between the mirror's reference body and the thin shell, a clever solution known as \emph{air damping}. 

Because higher frequency oscillatory modes are still present over the threshold provided by the air gap, additional damping methods must be provided to enable system frequencies over that threshold. MAPS addresses this by using a modified PID controller that includes a velocity-based damping control, which, if configured correctly, allows it to reach system frequencies of 1000 Hz.

However, the damping provided by the air gap has limitations. The outer three rings of actuators are close to the open edge of the air gap, and air damping becomes less effective at this edge. Therefore, resonance oscillation is still problematic in this region, and the mirror becomes hard to control in its outer regions. The velocity-based damping control was specifically designed to address this.

These vibrational modes have implications affecting positional control. Uncontrolled oscillation can be problematic, and resonance modes introduce another set of oscillations that are not controlled by the system. 

The interaction of this external oscillation with the PID's attempt at creating controlled oscillation can, on one hand, cause constructive interference, increasing the amplitude of resonance oscillation, and potentially destabalize the actuator's PID. On the other hand, destructive interference can damp the PID's action, hindering its ability to reach the set point. And further, this unwanted oscillation can interfere with the process of tuning itself (see Sec.~\ref{sec:PS}). This is the nature of the second aspect of the actuator control dilemma.


\section{THE MAPS ACTUATOR CONTROL SYSTEM}
\label{sec:ItMeth}

\subsection{Introduction}

The MAPS actuator control system was designed by systems engineer Keith Powell, whose doctoral thesis \emph{Next Generation Wavefront Controller for the MMT Adaptive Optics System: Algorithms and Techniques for Mitigating Dynamic Wavefront Aberrations} \citep{Powell12} contains almost the complete blueprint for the reasoning and design of the MAPS actuator control system. The design is the logical evolution of a lineage that started with the first ASM prototype of what was to become the MMT336, called the P30.

Although voice coil magnetic field actuators with capacitive position sensors were first introduced in the literature in 1993 \citep{salinari94}, it wasn't until the P30 was fabricated in 1998 that their use in the control of a thin shell deformable mirror was successfully demonstrated. Many of the control concepts that are still in use in MAPS and the Large Binocular Telescope adaptive optics system were developed for, and tested on, this miniature ASM. Because this developmental process was largely the result of a collaboration between Italian academic and private sector business, the control concepts are often referred to as the \emph{Italian Methodology}.

Any ASM that uses the Italian methodology has the following attributes:

\begin{itemize}[noitemsep]

    \item Voice-coil type non-rigid magnetic field generating actuators with built-in capacitive sensing.
    \item Magnetically levitated thin-shell mirror with magnets and a reference body, with a gap which allows:
    \item Air gap velocity damping.
    \item A PID-based controller, usually of the type PD\Plus$\!f$\Plus$\!\int$, (Proportional-Derivative controller with:
    \item Feed-forward loop and outer integrator loop.)
    \item Minimal use of the derivative function in recognition of positional noise.
    \item A system configuration of the SISO decentralized type.
\end{itemize}

Whereas ASMs of other design schools and their control systems may share some of these traits, all of them together are the hallmarks of this now dominate methodology. MAPS has inherited all of these attributes, and adds a few more. 

\subsection{MAPS Control System}

When discussing a control system, there are two aspects to consider. One is the overall structure of the system, and the other is the structure of the controller.

\subsubsection{Structure of the MAPS control system}
\label{sec:structure}

Because each of the MAPS actuator controllers has only one input and one output, the control system type is referred to as \emph{single input/single output}, or SISO. The job of the actuator controllers is to take an input position and generate an output coil current. Controllers dedicated to this type of task fall into the category of \emph{motion controllers}. Because the input to the controller is taken at the same spatial location on the controlled object that the output acts upon, it is further referred to as a \emph{coupled input-output pair}. 

A collection of SISO controllers each acting fully independently of each other is called a \emph{Fully Decentralized Control System}. The individual controllers in a decentralized system each control processes and variables that are completely decoupled from other controllers. 

ASMs push the above definition, however, as the controlled variables (position and output current) of any actuator are not independent of other actuators. Each actuator is effectively coupled to other actuators through the mirror. A system whose SISO controllers are not fully independent is sometimes referred to simply as a \emph{Decentralized Control System}. 

MAPS is therefore a \emph{decentralized, coupled input-output SISO control system}. Although beyond the scope of this paper, systems of this nature are prone to inherent oscillatory instability, for two reasons. First, coupled input-output pair SISOs that are tuned sub-optimally are prone to instabilities in their process variable (positional) calculations. Second, autonomous SISO systems that are coupled externally (via the mirror function in ASMs) are more easily destabalized because of the influence of one actuator's output on the input of a nearby actuator. Therefore, ASMs with ACS of this structure have an inherent tendency towards controller instability.

\subsubsection{MAPS controller}

The MAPS ACS, which is embedded in the onboard electronics of every actuator, is a modified PID controller incorporating a \emph{feed-forward loop} with a \emph{feed-forward bias} component. It differs from the Italian methodology in one important aspect: it does not have an outer integrator loop, but instead relies on the integrative term in the PID for this functionality. Integrative functionality is essential to the operation of an ASM because it is the correction term that protects against external influences to the mirror, such as wind and telescope vibration.

\begin{figure*}[!t]
    \vspace{2mm}
    \centering
        \includegraphics[height=2.9in]{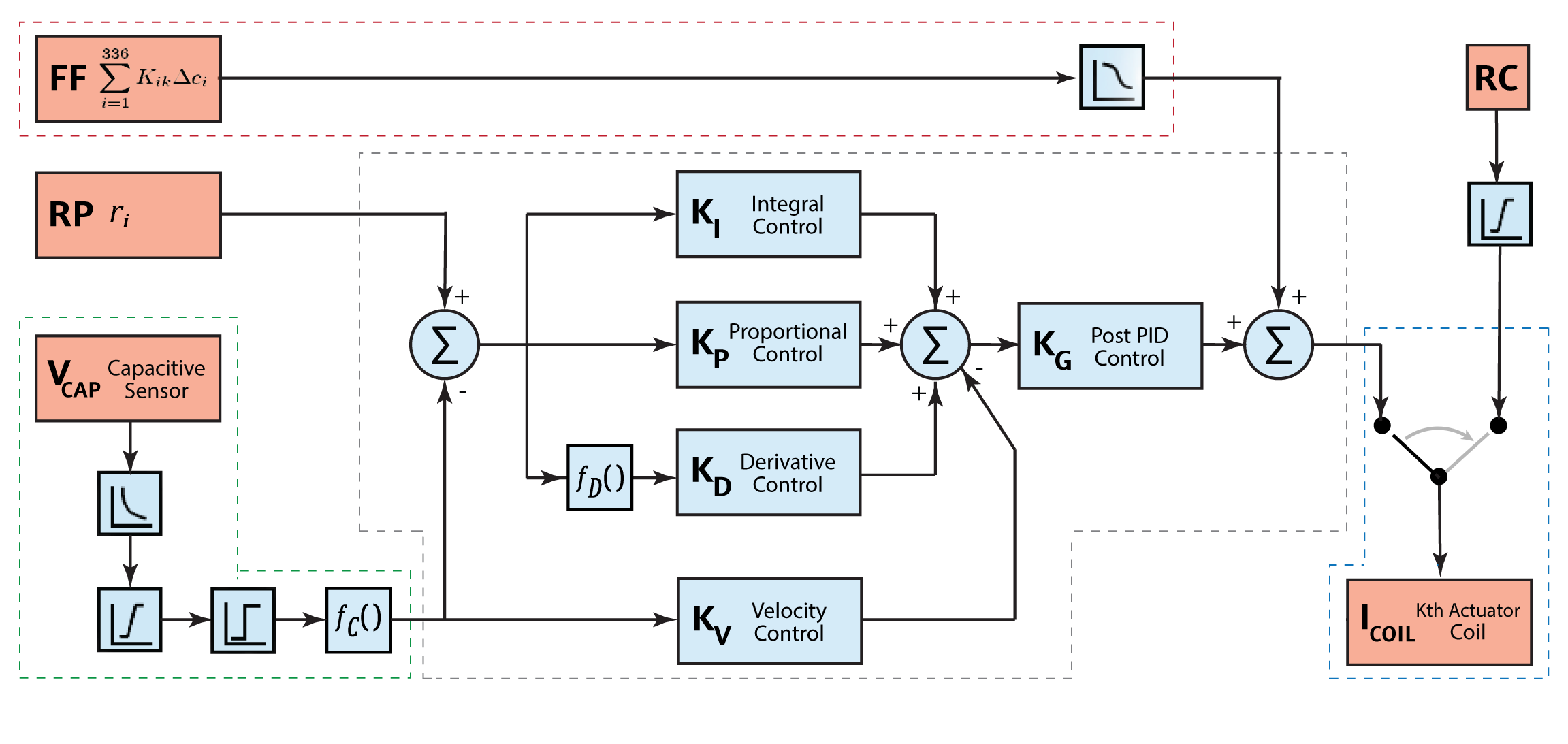}
    \caption{MAPS actuator control law block diagram. From \protect\citet{Downey19}.}
    \label{fig:MAPSCL}
    \vspace{7mm}
\end{figure*}

A few comments about feed-forward systems. Feed-forward in a control system is a method of estimating the PID output without waiting for the PID to cycle. It pushes the controlled object towards its set point by a known quantity before the first PID calculation, and then allows the PID to complete the process. This greatly reduces the settling time as well as the error, and the PID requires fewer oscillatory cycles to reach its set point. 

In the MAPS system, the primary feed-forward component is the \emph{feed-forward matrix}. This 336 by 336 matrix is basically a poke matrix, determined empirically during mirror characterization by poking each mirror segment and recording the response of all other mirror segments. In this way, it characterizes the \emph{mirror function}, which is the tendency of the mirror to resist any action that wants to bend its surface out of shape. By knowing the resisting force generated by pushing a mirror segment against its neighbors, an estimated current value required to move a segment to any displacement can be calculated and quickly applied.

MAPS also implements a feed-forward bias correction, which counteracts the bias magnet in each actuator that is a part of the safety system of the thin shell. When the ASM is in its rest state, the bias magnets protect against shell detachment. The feed-forward bias, applied with the feed-forward correction term, automatically negates this retaining force.

The MAPS control design incorporates feed-forward functionality as a required component, but incorporating it into the external control system code has been delayed to ensure that hard-coded safety mechanisms are in place, and the results discussed later in this paper do not incorporate its functionality. The effects of tuning values determined without feed-forward in place operating \emph{after} feed-forward is in use are unclear. Because the effect of feed-forward is to quickly reduce the error, we expect that the response of the PID should be similar to a small displacement without feed-forward. 

We  therefore expect that values determined by tuning without the feed forward loop will be applicable when feed forward is enabled. The end result is expected to be a substantially improved settling time over PID tuning alone.

\subsection{MAPS control law}

Any system controller is essentially a physical structure that houses and executes the system's control law. The control law is a complicated algorithm with user determined gain variables that establish the relationship between the input and the output of the control actuators, as well as establishing the timing and interaction of the outer and inner control loops. 

\subsubsection{Overview of the control law}

Figure~\ref{fig:MAPSCL} is a block diagram of the MAPS Actuator Control Law. It is divided into four principle components, indicated by dotted rectangles. Red indicates the feed forward component. Grey indicates the central PID. Green indicates the capacitive sensing component. Blue indicates the output. A key to the symbols and abbreviations in the diagram is given in Table~\ref{table:key}.

\begin{table}[b!]
    \centering
    \vspace{2mm}
    \caption{Key to the MAPS Control Law}
      \begin{tabular}{r l l r l l} 
      \toprule   
       Sym & Quantity & Description \\ [0.5ex]
      \midrule
       \multicolumn{3}{c}{\textit{Feed Forward}} \\ 
       \textbf{FF} & Feed Forward  & Calculated FF Values \\
       $\mathbf{K_{ik}}$ & FF Matrix & 336 x 336 Poke Matrix \\ 
       $\mathbf{\Delta C_i}$ & Delta Position & Error Values \\
       \multicolumn{3}{c}{\textit{Capacitive Sensing}} \\
       $f_D ()$ & Diff Filters & Impulse Response \\
       $\mathbf{V_{\text{CAP}}}$ & Segment Positions & Vector, 336 Pos Values  \\
       $f_C ()$ & Cap Filters & Pulse Time Shaping \\
       \multicolumn{3}{c}{\textit{Central PID}} \\
       $K_I$ & Integral Variable & Integration Gain \\
       $K_P$ & Proportional Variable & Proportional Gain \\
       $K_D$ & Derivative Variable & Derivative Gain \\
       $K_V$ & Velocity Variable & Mirror Damping Gain \\
       $K_G$ & Gain/Switch Variable & PID Switch \& Gain \\
       \multicolumn{3}{c}{\textit{Mirror Command}} \\ 
       \textbf{RP} & Run Position & Command Position \\ 
       RC & Run Coil & Operator override \\
       $r_i$ & ith actuator &  Command Position \\
       \multicolumn{3}{c}{\textit{Output}} \\	   
       $I_{COIL}$ & Coil Current & Controller Output \\
      \bottomrule
    \end{tabular}
    \label{table:key}
\end{table}

The core of the controller is a parallel-form PID, with $K_I$, $K_P$ and $K_D$ representing the standard integral, proportional and derivative gain terms. In addition to these standard terms is the \emph{Velocity damping function} with its tunable gain term $K_V$, and a \emph{Post PID Gain} Variable, $K_G$. $f_D ()$, the differentiation finite impulse response filter, is an additional set of variables for the derivative process. 

Outside the core is the capacitive sensing position system, with its pulse shaping $f_C ()$ settings, the feed forward loop with feed forward bias and feed forward smoothing, and the Run Coil operator controller override $RC$. The next subsection explain a few of these quantities.

\begin{figure*}[!t]
    \centering
        \includegraphics[height=1.5in]{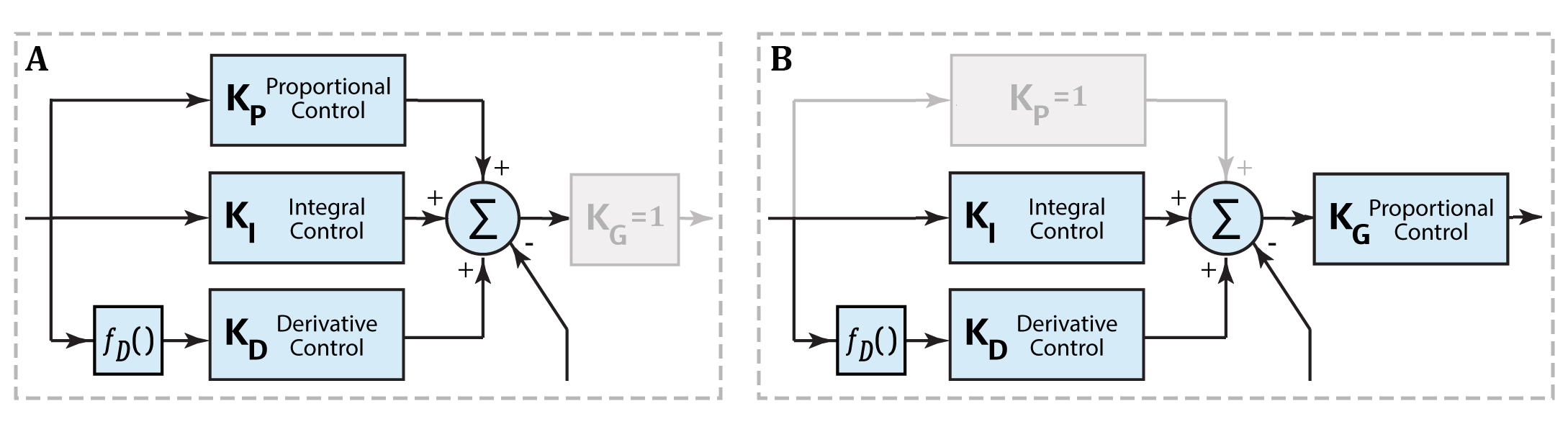}
    \caption{Effect of the $K_G$ variable. A) With $K_G=1$, the MAPS PID is a parallel form controller. B) Setting $K_P=1$ and using $K_G$ as a scaled gain variable effectively switches the PID to a standard form PID.}
    \label{fig:PIDform}
    \vspace{7mm}
\end{figure*}

\subsubsection{New control quantities}
\label{subsec:Mods}

The new controller implements a few new tuning quantities, discussed here in greater detail:

\textbf{Velocity damping $\mathbf(K_V)$}: The velocity damping variable is a setting designed to compensate for resonance modes that occur at operational frequencies over 500 Hz and to correct for the reduced air gap damping in the outermost actuator rings. Although unclear in the control law diagram, Powell's notes on his MAPS simulator (more on this in Section~\ref{sec:tuning}) shows that it follows a differentiation filter in the capacitive sensing system. Presumably, then, the filter is determining velocity from successive positional readings, and $K_V$ is a proportional gain term that scales the result. Due to the fact that this correction is subtracted from the overall calculation at the controller's second summation point, it acts to reduce the magnitude of the calculated output current.

\textbf{Run Coil ($\mathbf{RC}$)}: The Run Coil component of the controller is an open-loop override that allows direct actuator current values to be entered by the operator, essentially bypassing the controller. When the operator enters a value, the switch (indicated by the curved grey arrow in Figure~\ref{fig:MAPSCL}) automatically routes the value directly to the actuator. 

\textbf{Capacitive, Differentiation Filters ($\mathbf{f_C ()},\mathbf{f_D ()}$}): The \emph{capacitor finite impulse response filter} allows shaping of the 'Go Pulse' voltage signal that is read by all actuators to determine position. We have not begun working with this setting; it is currently set to default. The \emph{differentiation finite impulse response filter} is a time shaping control for the PID's derivative gain. Similarly, this is set at default.

\textbf{Post PID Gain ($\mathbf{K_G}$)}: This final term in the central controller has two purposes. The first is to convert the positional values that are the output of the rest of the central controller to current units (the feed forward component's output is already calculated in current units). The second, more interesting use of this variable is that it allows the core PID controller, initially in parallel-form, to be switched to standard-form, as shown in Figure~\ref{fig:PIDform}. The process of tuning for a standard-form PID is different then the process utilized here.

\begin{figure*}[!t]
    \centering
        \includegraphics[height=3.0in]{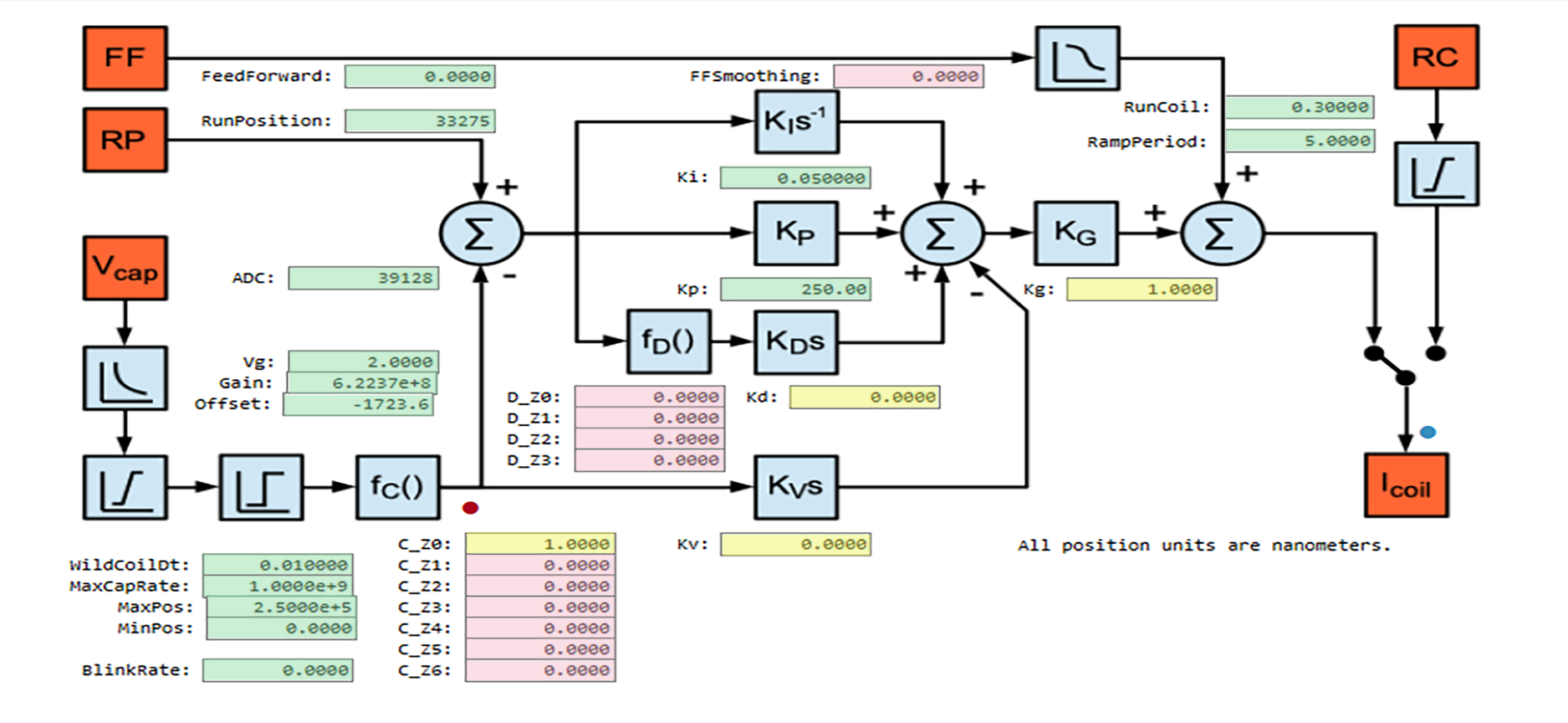}
    \caption{The operator interface for the MAPS actuator controller. Each rectangle represents a setting. Rectangles in green represent quantities used in the first round of tuning experiments. Rectangles in yellow represent quantities that we understand in concept, but have been left to a second round of tuning experiments and are currently set to default values. Rectangles in red are quantities of which we have limited understanding.}
    \label{fig:FullLaw}
    \vspace{5mm}
\end{figure*}

\subsubsection{A glimpse at full tuning}

The above discussion has highlighted the most important details about the MAPS actuator controller. There are more settings in MAPS then just the ones emphasized here, though. The complete interface for the actuator controller is shown in Figure~\ref{fig:FullLaw}. 

Each rectangle in the image is either a variable setting or output value. The rectangles are color coded to represent the current state of our understanding. Rectangles in green are variables we fully understand, and are the quantities we've been working with in this first round of tuning. Rectangles shaded in yellow are quantities that we conceptually understand, but have been left for the next series of experiments and are set at default values. And rectangles shaded in red are quantities set at default values that we have limited understanding of their correct usage. 

\subsubsection{Sequence of operations}
\label{sec:flow}

As a final step in understanding the MAPS controller, we list its sequence of operations. There are two control loops running simultaneously. The \emph{outer loop} is the traditional adaptive optics process in which the wavefront sensor determines the shape of the incoming wavefront and the reconstructor calculates the phase conjugate needed to correct it. The  \emph{inner loop} is the PID control loop.

The inner loop is required to run at a higher frequency then the outer loop, so that the inherently time consuming process of getting actuators to settle to their setpoint positions can complete before the next cycle of the outer loop. For MAPS, the design specifies that the outer loop maximum frequency is 1 \si{kHz}, and that the inner loop cycles at 100 kHz, giving 100 cycles of the inner loop for every one cycle of the outer loop. All of the actions listed below, then, take place in outer loop cycles of 1 ms length. During each 1 ms cycle, 100 cycles of 10 $\mu \text{s}$ inner loop passes occur. Actuators are being queried, sent commands, and move to positions 100,000 times per second. 

\vspace{2mm}

\noindent\textbf{1) Outer Loop Cycle:} At the start of each outer loop cycle, the controller receives a vector of commanded mirror positions from the AO system. This is the controller input $\mathbf{r}$, or `\textbf{RP}', and triggers the inner loop to start cycling to position the actuators. 

\vspace{2mm}

\noindent\textbf{2) Inner Loop Cycle:} The inner loop starts, and repeats, following this sequence: 
\vspace{-2mm}

\begin{figure*}[!t]
    \centering
        \includegraphics[height=3.2in]{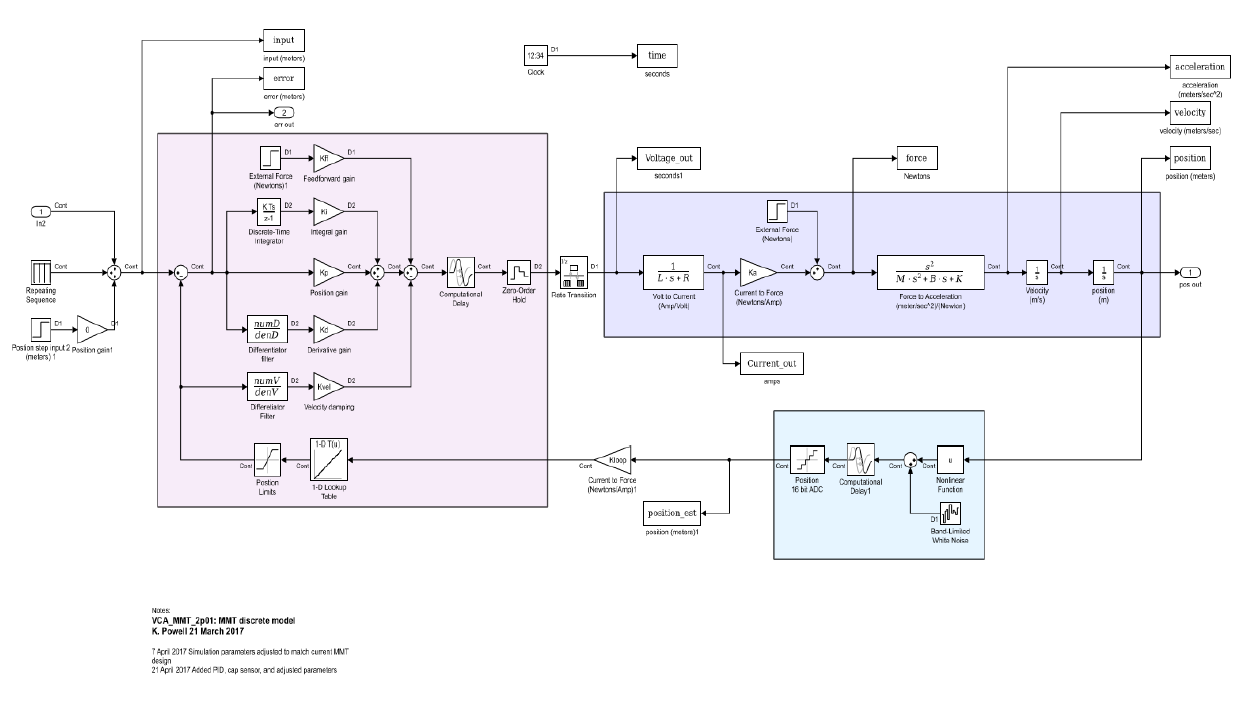}
    \caption{A control diagram representing a portion of the SPAM MAPS simulator. Reconstructing the simulator is being considered as a model-based tuning tool. From \protect\citet{powell17}.}
    \label{fig:SPAM}
    \vspace{7mm}
\end{figure*}
\vspace{5mm}

\begin{enumerate}
    \item In the first cycle of the inner loop, the controller receives a vector of current mirror positions as determined by the capacitive sensing system. This is input column vector ${\mathbf{V}}_{\text{CAP}}$. As part of this process, the system determines whether the CapSense input from any actuator is exhibiting rapid oscillation, or WildCoil. If it does, the system throws a `WildCoil' alert to the operator, and, depending on safety settings in the system's housekeeping system, may also safe the mirror, shutting the process down. Also, position values fed out by the CapSense system are checked to see if they are exceeding the system's maximum and minimum positions. The capacitive filter values $\text{f}_{\text{C}}(z)$ are then applied, and the final measured mirror position vector ${\mathbf{V}}_f$ is sent to the first branch point:
    $$ \mathbf{V}_f = f_C(z) {\mathbf{V}}_{\text{CAP}}. $$

    \item At the branch point the signal path splits. The first branch of the split goes to the summation point, where ${\mathbf{V}}_f$ is subtracted from the incoming $\mathbf{r}$ vector, creating the delta position (error) vector $ \mathbf{\Delta c}$:
    $$ \mathbf{\Delta c}  = \textbf{r} -\mathbf{V}_f.$$
    \hspace{-1mm} These are the \emph{positional change} values that will eventually be converted by the controller to coil voltages to move the mirror to the position commanded by the outer loop.

    \item Although not explicitly shown on the diagram due to clarity considerations, $ \mathbf{\Delta c}$ is also sent to the feed-forward loop, where it is acted on by the feed-forward matrix to produce the feed-forward vector. The feed-forward bias is added to this, creating the \emph{feed forward term}:
    $$\textbf{ff}  = \mathbf{\Delta c}_i  \mathbf{K}_{ij} + \mathbf{f}_B.$$

    \item Next, $\mathbf{\Delta c}$  is sent through the PID controller. To do this, its path is split in three. It is important to realize that \emph{splitting the path does not split the signal}; each path carries the same signal. Each PID variable acts on the same $\mathbf{\Delta c}$. The proportional gain process is applied to the signal:
    $$\mathbf{P} = \text{K}_P \mathbf{\Delta c}.$$
    The integral gain process integrates the cumulative error and applies the integral gain value. Note that $\tau$ is used here as the variable of integration, and the bounds are run from 0 to t. This indicates that the cumulative error from t=0 to the present moment is integrated:
    $$\mathbf{I} = \text{K}_{\text{I}} \int_{0}^{t} \mathbf{\Delta c}(\tau) d\tau.$$
    The derivative process takes the time derivative of the error and applies the derivative gain and filter values: 
    $$\mathbf{D} = f_d() K_D \frac{d}{dt} \mathbf{\Delta c}.$$

    \item At the second summation node, these values are all added together. If a value other than zero has been entered into the velocity damping loop, activating it, then $\mathbf{V}_f$ is acted on by $K_V$:
    $$\mathbf{V} = K_V \mathbf{V}_f.$$
    
    \item The result of the summation node is then:
    $$\mathbf{\Sigma} = \mathbf{P} + \mathbf{I} + \mathbf{D} - \mathbf{V}$$

    \item At the third summation node, the $K_G$ gain is applied to this, converting positional commands to current values, and then summed with the feedforward term: 
    $$ \mathbf{I}_{COIL} = K_G \mathbf{\Sigma} + \mathbf{FF} $$

    \item  This final value is then sent to the actuator coil, and the inner loop repeats from step one of the inner loop cycle.

\end{enumerate}

\vspace{4mm}
The only exception to the above is if the operator has chosen one or more specific actuators and entered RunCoil value(s) for them. This bypasses the actuator's controller output and sends the entered value directly to the coil.


\vspace{3mm}

\section{TUNING THE MAPS ASM}
\label{sec:tuning}

This whole endeavor is about \emph{`tuning the ASM'}. So let's  define more precisely what that means. The standard definition of `tuning' a controller is:

\blockquote{...the process of determining the controller parameters which produce the desired output... [it] allows for optimization of a process and minimizes the error between the variable of the process and its set point. 

When a mathematical model of a system is available, the parameters of the controller can be explicitly determined. However, when a mathematical model is unavailable, the parameters must be determined experimentally. \citep{website:bennett}}

One of the things that this definition hints at is the use of a mathematical model to determine the appropriate controller parameters. 

Indeed, this seems to be the preferred way to tune complex control systems with many actuators and fragile controlled objects, at least in the industrial world. Doing so allows one to experiment with values for control gains without the possibility of unintended consequences.

Several authors, however, have written about the difficulty of modeling adaptive secondary mirror systems, due to mirror resonances, actuator coupling, and the stability issues with SISO systems.  Intriguingly, however, there apparently was such a model for the MAPS ASM, called \emph{SPAM}.

SPAM stands for \emph{SPAM: the PAlindromic Maps simulator}, and was a MATLAB code designed to simulate the operation of the ASM. It was hinted at in a presentation Kieth Powell gave in 2017, but unfortunately seems to have been lost. However, the general control diagram representing the code does, and is shown in Figure~\ref{fig:SPAM}. 

We are considering whether using the diagram as a basis for coding a simulator may be useful to provide a model based method of tuning. This is future work, however, and for now, MAPS must be tuned experimentally.

\subsection{Initial Performance}

When the first MAPS engineering runs took place, the ASM was effectively running on a limited set of default tuning values. The ASM was also exhibiting a few problematic issues.

\subsubsection{Running on defaults}

All actuators were set to the same values, as follows:

\begin{itemize}[noitemsep]
  \item \textbf{Proportional Gain}: $K_P=100.00$
  \item \textbf{Integral Gain}: $K_I=0.05$
  \item \textbf{Derivative Gain}: $K_D=0.00$
  \item \textbf{Velocity Gain}: $K_V=0.00$
  \item \textbf{Post Gain}: $K_G=1.00$
\end{itemize}

It was not known whether these tuning values were appropriately set, or even whether or not they represented `safe' values (providing minimal but non-destructive actuator response). 

\subsubsection{Actuator failure}

At start, the ASM went on sky with twelve actuators disabled due to issues with their electronics. Actuators can fail due to a number of reasons, from non-responsive electronics to bad capacitive sensing systems. MAPS classifies these errors into several different types, as follows:

\begin{itemize}[noitemsep]
  \item \textbf{DEADBEEF}: Actuator has failed internal checksum test; halts system start. Fatal error;
  \item \textbf{DEADFEED}: Actuator has electrically failed. Fatal error;
  \item \textbf{CAPREAD}:	 Actuator's capacitor's ADC has failed. Fatal Error;
  \item \textbf{NOVAL}:    Actuator payload is all zeroes. Potentially fatal error if actuator can't reset itself;
  \item \textbf{EXCEEDS}:	 Actuator persistently attempts to position itself beyond the system limitation for a single step. Non-fatal but annoying; 
  \item \textbf{BADCAP}:   Actuator is displaying bad positional information. Non-fatal but potentially serious.
\end{itemize}

The ASM went live with 7 DEADFEED actuators, 4 DEADBEEF actuators, and 1 CAPREAD actuator. 

\subsubsection{Actuator Noise}

In addition, the ASM started life with with excessive levels of actuator positional noise. Actuators can become noisy for several reasons. Ground loops in the power supply contribute heavily to actuator noise, as do ground faults in the ASM central hub electronics. Background electric fields from heavy equipment interacting with the conductive surface of the underside of the thin shell can make the Go pulse voltage signal dirty, leading to errant positional values. We found that noise had dependence on telescope altitude, which reflects the shifting of the orientation of the secondary as it moves vertically.

The MAPS design specification indicates an RMS noise value of \SI{\pm5}{\micro\metre}. The initial noise characterization, carried out before the first engineering run, found an RMS level of \SI{\pm22}{\micro\metre}, with individual actuators ranging as high as \SI{\pm50}{\micro\metre}. 

\begin{figure*}[!t]
    \centering
        \includegraphics[height=1.95in]{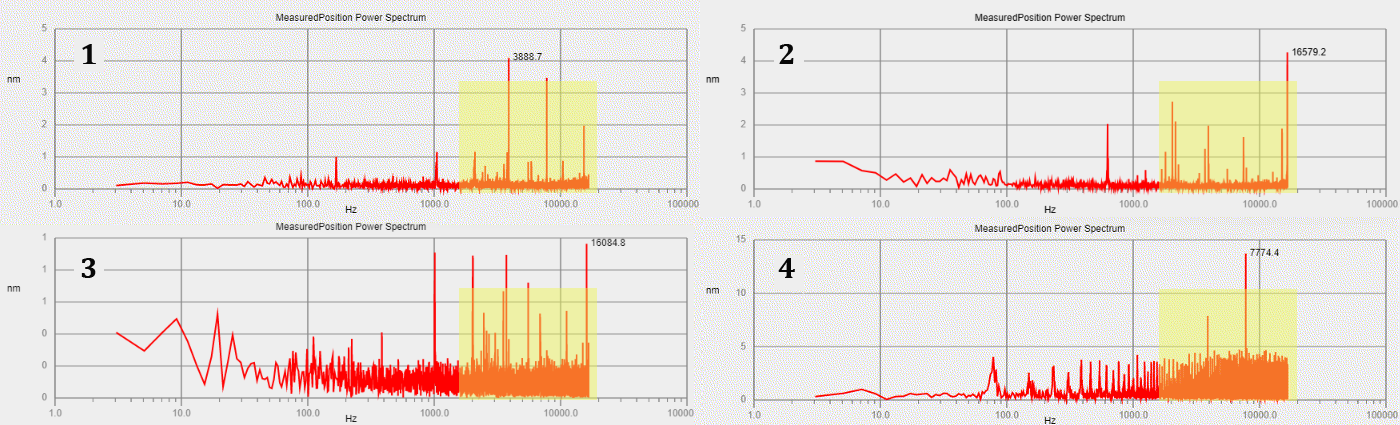}
    \caption{Four frames taken while tuning an unconstrained actuator. Sequence goes from top left to top right to bottom left to bottom right.}
    \label{fig:FFT04}
    \vspace{7mm}
\end{figure*}

\subsubsection{Initial Performance}

The ASM performed poorly under these conditions. The mirror continuously safed itself, the result of large numbers of individual actuators independently triggering WildCoil alerts (see Section~\ref{subsubsec:WildCoil}.) Actuators exhibited power consumption at levels considerably over design specification, and actuators overheated, requiring frequent mirror down time for cooling. The system was limited to operating frequencies of less than $\SI{100}{\hertz}$.  

Suspecting that the default actuator gain values may be partly responsible for these issues, the first round of experimental tuning was undertaken. As tuning under conditions of elevated noise levels and disabled actuators was expected to yield problematic results, considerable effort went into diagnosing actuator failure and to find the cause of the elevated noise, with little success. 

\subsection{Tuning of the proportional gain}

After several engineering runs, we decided to begin by focusing on actuator proportional gain (PG); this is the primary tuning variable controlling the PID oscillatory process. The proportional term of the controller's correction is simply the product of the error and the PG value. The larger the PG, the larger the acceleration the mirror segment is subject to. If the value is set too high, the mirror is kicked too hard, and the resulting amplitude of the oscillation is large, requiring more time to settle to its set point, or worse. If the PG value is set too low, oscillation can't start. The logical result of this is that the tuning value is set correctly at the point where the mirror segment just begins to oscillate.

The reason that elevated actuator noise is problematic to the process of tuning is that tuning is fundamentally the adjustment of an oscillatory behavior, and actuator noise represents yet another kind of random oscillation. If we wish to discern the point at which changing the proportional variable results in oscillation, the existence of pre-existing random oscillation interferes, as does mirror resonance oscillation.

\subsubsection{Tuning by power spectra}
\label{sec:PS}

How do we determine when the mirror has just begun to oscillate? Elwood Downey, the software engineer of MAPS, left us with a powerful set of tools. The MAPS interface includes two real-time plots for every actuator controller: a position vs. time (PVT) plot, and a live Fast Fourier Transform of the PVT plot. The FFT plot essentially shows the power spectrum of mirror oscillation.

In addition, MAPS provides a tuning mode, which samples the mirror position at a faster rate then during normal operation, which allows the FFT to show frequencies in the power spectrum up to 30 kHz.

The procedure for tuning a single actuator's $K_P$ value is as follows. Starting with an actuator with its PG variable set to minimum, i.e., $K_P = 1$, we load a system configuration file, load a flat, and then set the system to tuning mode and display the actuator's PS plot. We take careful note of what the power spectrum looks like, and then begin to increase $K_P$ in minimal steps, maybe 10 points at a time, and watch the PS plot. The goal is to find the point that the already existing oscillatory behavior begins to change. 

There are two types of distinct changes in the power spectrum that we find by doing this, depending on the actuator's physical position. 

\subsubsection{Type I oscillators}

The first type of behavior, which we call Type I, is observed in actuators that have little or no constraining factors (\emph{unconstrained} actuators). Actuators in the outer three rings are undamped (or less damped) by the air gap, and exhibit oscillation occurring over a broad frequency range, as can be seen in the sequence in Figure~\ref{fig:FFT04}. 

For these actuators, increasing $K_P$ leads to oscillatory behavior across a large range of frequencies, generally starting around 1 kHz and extending to the end of the spectrum. The effect happens quickly at a threshold value of proportional gain, when the power spectrum rapidly broadens after a small increase in the PG. We take note of this threshold value, and use this value as the value of $K_P$.

\begin{figure*}[!t]
    \centering
        \includegraphics[height=1.95in]{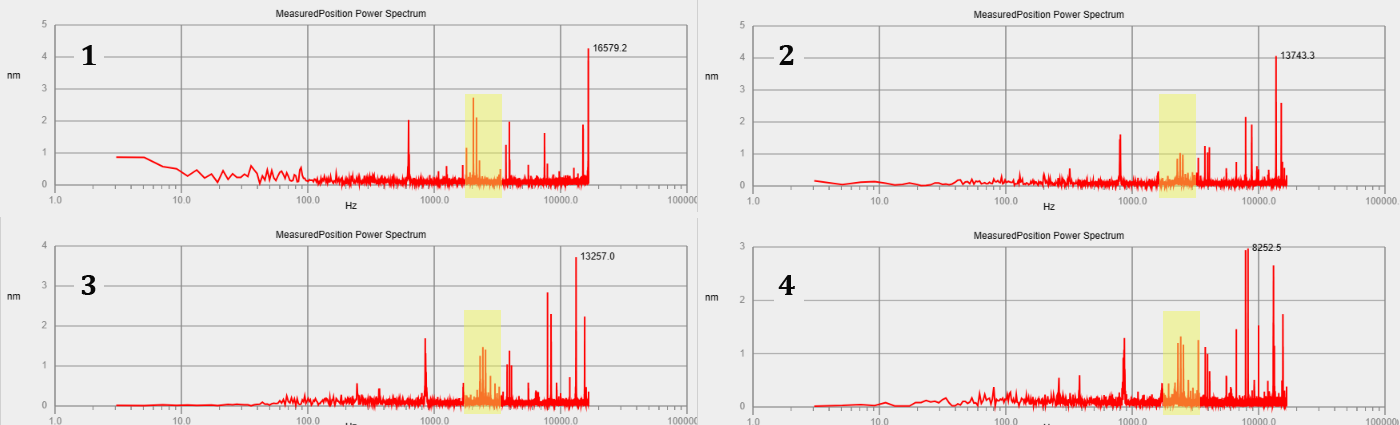}
    \caption{Four frames taken while tuning a constrained actuator. Sequence goes from top left to top right to bottom left to bottom right.}
    \label{fig:FFT03}
    \vspace{7mm}
\end{figure*}

\subsubsection{Type II oscillators}

The second type, which we call Type II, is shown in the sequence in Figure~\ref{fig:FFT03}, and is typical of a \emph{constrained} actuator. 

Constrained actuators are actuators in the center rings (rings 1-8), or actuators in the outer rings (rings 9-11) that are situated next to an actuator that has been disabled. These actuators are either damped by the air gap (inner rings) or by the coupling influence of the inactive actuators next to them (outer rings). 

When we start increasing the value of the PG, we find that a particular region of the power spectrum begins to develop tightly packed peaks, as can be seen in the yellow highlighted areas. The effect is subtle at small increases, but can easily be pushed to the point where the area grows considerably. We call this effect the `2K Forest', because this form of oscillation generally centers itself around 2 kHz on the PS. Since we want the point that oscillatory behavior begins, we take note of the PG value that triggers this response, and this becomes $K_P$.

\subsubsection{First tuning configuration results}

By repeating this process for every active actuator, a process which takes roughly 15 minutes per actuator, we completed the first MAPS tuning configuration, and used it during the next engineering run. The immediate results were obvious in several ways. 

First, the incidence of WildCoil events was drastically reduced. Actuator power consumption was also reduced, as were overheating events. However, rudimentary determination of system latency, a reflection of mirror latency, did not show significant improvement in settling time. Further, the maximum system operational frequency remained capped at 100 Hz. It was not immediately clear, however, as to whether this was a result of the tuning configuration, or was due to myriad other `new system' issues.

\subsubsection{ASM rebuild}
\label{sec:rebuild}

After the first series of engineering runs, ending with the first application of the tuning configuration described above, the MAPS ASM went into the laboratory clean room to be serviced, to address the noise issue and the disabled actuators. By this time, more actuators had failed, giving a total of 35 disabled actuators, and a large number of actuators were exhibiting high levels of noise, 35 of which were consistently triggering WildCoil events.

During the rebuild, multiples issues were found affecting actuators. Several DEADBEEF actuators, for example, were found to have cable or connector issues. Many actuators were either not aligned co-axially with their bore holes, or were found to be at the wrong distance from their mirror magnet. During the ASM's period on the lab, actuators that were defective were replaced; the alignment of all actuators was checked  and corrected, electrical connections were reseated, and cable was tested and replaced. 

\begin{figure*}[!t]
    \centering
        \includegraphics[height=3.2in]{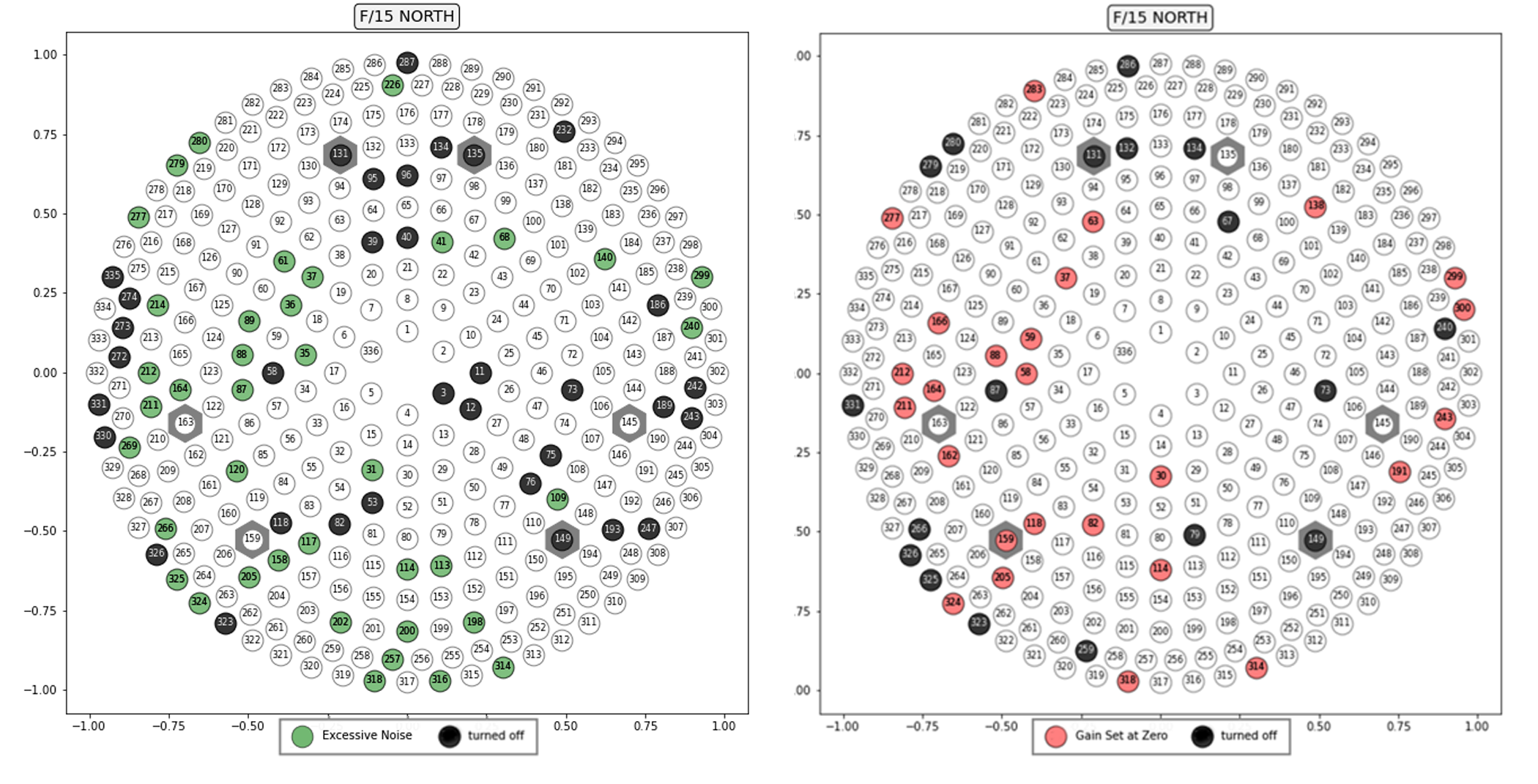}
    \caption{Plots of the MAPS actuators, before (left) and after (right) the ASM rebuild. Black indicates a disabled actuator. Green indicates an actuator with a high noise level. Red indicates a 'floating' actuator with its CSS system disabled. }
    \label{fig:Rebuild}
    \vspace{7mm}
\end{figure*}

As a result of this down period, the ASM went back to the mountain in substantially better operational condition. Preliminary noise characterization results of between \SI{5}{\micro\metre} and \SI{8}{\micro\metre} RMS indicated that positional noise had been reduced almost to its design specification. Disabled actuators had been reduced from 35 to 18; and functional actuators went from 262 (78\%) before the rebuild to 319 (95\%) after. Figure~\ref{fig:Rebuild} shows the pre- and post-rebuild actuator maps.

\subsection{Attempts at Ziegler-Nichols}

The next substantial effort towards tuning came after the rebuild. We decided to attempt to implement Ziegler-Nichols (ZN) tuning methodology, using the FFT technique described in Section~\ref{sec:PS}. ZN tuning is by far the most commonly used parallel-form PID tuning method, because of its intuitive procedure, relative simplicity of application, and its applicability to most PID situations. It is essentially heuristic, which goes a long way to explaining its simplicity and popularity. This is the method control engineers will typically use before attempting more difficult model-based methods. 

We went through the 336 actuators again, applying the PS process to determine the actuator's oscillation frequency. The values we gained by doing this were inconsistent with those found during initial proportional gain tuning, as had been expected, due to the reconfiguration of the ASM. 

\subsubsection{Description of Ziegler-Nichols}
\label{sec:ZNDescrip}

In ZN, we utilize these power spectra results to determine the value of the P, I, and D variables. ZN is essentially a series of empirically determined equations that lead to optimal control tuning. For MAPS, the process goes like this:

\begin{enumerate}[noitemsep]
    \item Set $K_I$ and $K_D$ to zero.
    \item Set $K_P$ to 1.0.
    \item Increase $K_P$ until the loop output begins to oscillate. The gain value that causes this becomes $K_U$, known as the `ultimate' frequency.
    \item The frequency at which the oscillation occurs becomes the value of $T_U$.
    \item Using these values, set the values of $K_P$, $K_I$, and $K_D$ as shown below:

$$ K_P = 0.60 K_U $$
$$ K_I = 1.2 \frac{K_U}{T_U} $$
$$ K_D = 3 \frac{K_U T_U}{40} $$

\end{enumerate}

These, then, are the gain values that are used to calculate the PID correction:

$$ u(t) = K_P e(t) + K_I \int_{0}^{t} e(\tau) d\tau + K_D \frac{de(t)}{dt} $$
 
The method is relatively straightforward, at least for most controllers, which is why the method is so commonly used. But with the MAPS controller, the method failed.

\subsubsection{Failure of Zeigler-Nichols}

The failure of Ziegler-Nichols was immediately obvious, as the values that it produced for the Integral and Derivative gain values were not reasonable. As an example, consider the tuning of actuator 125, shown in Figure~\ref{fig:FFT03}.

At the point that we determined oscillation had clearly begun (lower left plot), we had increased the proportional variable until $K_P = 250$, so $K_U = 250$. The central frequency of the  oscillation is almost exactly \SI{2}{\kilo\hertz}, which gives a period value of \SI{0.5}{\milli\second}. We use this value then to find what our $K_P$, $K_I$ and $K_D$ values should be: 

$$K_P = 0.60K_U = 150$$
$$ K_I = 1.2 \left( \frac{250}{0.5 \si{ms}} \right) = 6 \:\text{x} 10^{5} $$
$$ K_D = 3 \left( \frac{(250)(0.5 \si{ms})}{40} \right) = 9.4 \:\text{x} 10^{-3}$$

\begin{figure*}[!t]
    \centering
        \includegraphics[height=3in]{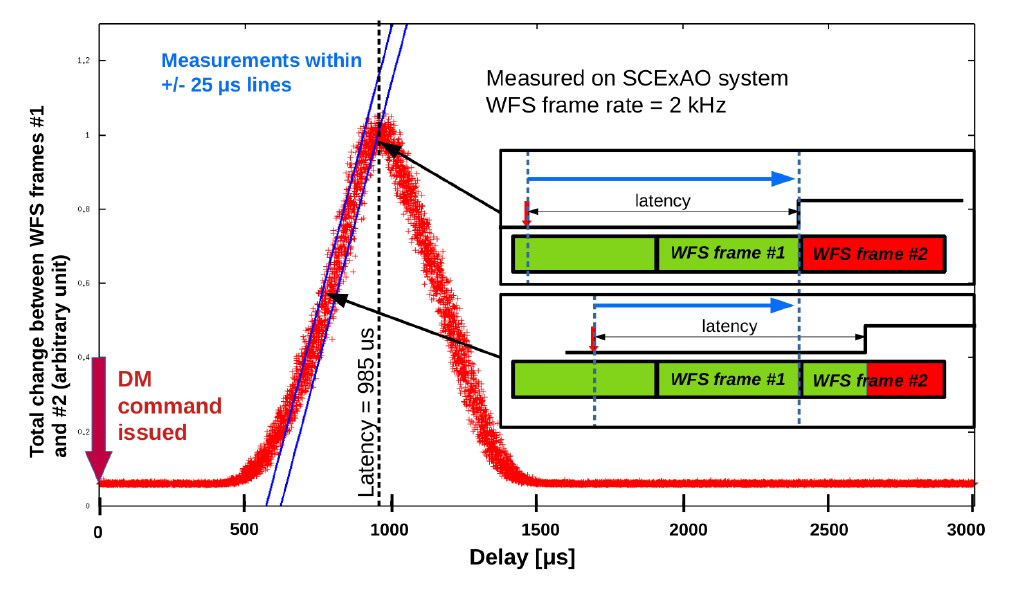}
    \caption{Mirror Latency (MLAT) measurement. This measurement was taken on the SCExAO sytem with a Pyramid WFS and a MEMS type deformable mirror, and shows a measured hardware latency of \SI{985}{\micro\second}. From \protect\citet{guyon18}.}
    \label{fig:MLAT}
    \vspace{5mm}
\end{figure*}

A glance at the integral value shows that it cannot be correct, as can be seen by simple inspection.

Let's say that the system's first pass through the controller starts with the CSS determining that a particular actuator's mirror segment has a positional error of \SI{5}{\micro\metre} from its set point. Let's apply the error correction calculation using these PID variables with $\mathbf{e}_{125} = \SI{5}{\micro\metre}$.

The first term, $K_P e(t)_{125}$ gives $250 (\SI{5}{\micro\metre}) = \SI{1.25}{\milli\metre}$. Although this seems large, this is pretty typical value for the initial proportional term. The second term takes the sum of the error until the present moment, which, in the first cycle, is \SI{5}{\micro\metre}, and multiplies it by the integral variable, giving us $\num{6e5} \SI{5}{\micro\metre} = \SI{3}{\metre}$. This is the type of unreasonable value mentioned previously. There is no third term, as the error is neither decreasing or increasing on the first pass, but this is enough to show the method fails. 

The integral term will immediately attempt to push the mirror too far. And on every subsequent pass, it will only explode further as the original pass has now driven the error into a value a thousand times too large. 

This exercise, along with the experience of attempting this method for hundreds of actuators, has shown that ZN cannot possibly work according to its cookbook recipe. 

There are three potential reasons for this. First, a glance at the upper left image in Figure~\ref{fig:FFT03} shows an actuator in its resting state with a minimal proportional gain variable set. Even in this state, the FFT plot shows oscillatory behavior at several frequencies, more than likely the result of positional noise or exterior vibration. This behavior can mask the start of oscillation when increasing the proportional gain, and all actuators display this behavior to differing degrees. 

Second, the inter-actuator coupling caused by the mirror function may be damping the range of frequencies in which the mirror begins to oscillate. And third, the constructive or destructive interference caused by mirror resonances described in Section~\ref{sec:resonances} may be interfering with the mirrors oscillation in response to increasing the proportional gain. 

It would appear that Ziegler-Nichols is not appropriate for tuning the ASM's actuator control system. 

\subsection{Tuning by ring}

\subsubsection{Modified tuning sets}

A further attempt at creating tuning configurations consisted of taking the values of the proportional gain determined in the previous PS experiments and modifying them by applying a multiplier evenly across all actuators, modified only by the determination that a specific actuator's gain should be kept at its current value, usually the result of a particular actuator being noted as excessively noisy or producing erratic positional output. 

In this way, three more tuning sets were created: MinTune (x2), ModTune (x5) and MaxTune (x10), with the multiplier affecting only the proportional variable. Note that all attempts at tuning to this point were modifications to the proportional gain only. Other tuning variables remained at default. For an unknown system, the optimal value of the proportional gain should be determined first and independent of modifications to other gain values. This has, indeed, been shown by experiments where modifying multiple variables in the same tuning set proved ineffective or counterproductive.

Also, With the MAPS ASM, we have a good understanding of the range of magnitudes we can use for the proportional gain value; we have no such understanding of the range of magnitudes acceptable for the integral and derivative values. We assume the original default value for the integral gain is reasonable and safe. 

\subsubsection{Ring tuning}

One of the important concepts that one eventually stumbles upon in reading the literature is the idea of \emph{ring tuning}. At the start of this process, we viewed actuator tuning as a single-actuator-at-a-time process. As each actuator takes approximately fifteen minutes to tune, the required time becomes on the order of weeks. The seeming necessity of this approach lay in its results: it appeared that many actuators displayed a different result when their controller's $K_P$ value were adjusted. Seeing this behavior gave credence to the idea that every actuator needed to be individually adjusted.

But mathematically, this is not the case, and the Circulant Matrix mathematics presented in Grocott's work \citep{grocott97} makes this clear. The fundamental value, indeed the only value, which differentiates an actuator's tuning is \emph{their distance from the center of the mirror}. Because of this, all actuators in any given ring can be tuned to the same values, which, as can be immediately seen, is a tremendous time saver when working out tuning regimes, dropping the number of tuning attempts from 336 to 12 and the time required from weeks to hours.

\begin{table}[t!]
    \centering
    \caption{First group tuning configuration files}
    \vspace*{3mm}
      \begin{tabular}{c l c l} 
      \toprule
      Num & Name & Gain & Notes \\ [0.5ex]
      \midrule
       01 & CleanedCurrent &  & Operational Conf \\
       02 & ElwoodDefault &  & Initial default \\
       03 & MinTune & P & Ind PGs x2 \\
       04 & ModTune & P & Ind PGs x5\\
       05 & ModTune\_IntR & P,I & 03 w IGs red \\
       06 & ModTune\_IntI & P,I & 03 w IGs inc  \\
       07 & MaxTune & P & Ind PGs x10  \\
       08 & Ring\_MinTune & P & Ring, PGs from 02 \\
       09 & Ring\_ModTune & P & Ring, PGs from 03  \\
       10 & Ring\_ModTune\_IntR & P,I & 08 w IGs red  \\
       11 & Ring\_ModTune\_IntI & P,I & 08 w IGs inc  \\
       12 & Ring\_MaxTune & P & Ring, PGs from 06  \\
      \bottomrule
    \end{tabular}
    \label{table:groupone}
    \vspace{5mm}
\end{table} 

\subsubsection{Disabled actuators and coupling effects}

This leads to an obvious question: why do MAPS actuators in the same ring behave differently when they should behave the same? If all actuators are exact functional copies of each other, and assuming that the design of the actuator is sound, there is something else happening. And there is: the presence of disabled actuators.

Actuators that are disabled put an increased load on the actuators that are functional, changing the inter-actuator coupling geometry. And actuators that are dysfunctional, usually with their capacitive sensing function disabled (`floaters'), create holes in the collocated SISO scheme, causing those actuators to exhibit anomalous oscillatory behavior. 

This leads us to an important conclusion: tuning can only be as good as the number of fully functioning actuators. The more disabled or faulty actuators there are, the more the tuning process becomes degraded by non-functioning actuators effect on actuator coupling.

\subsection{MAPS Tuning Test Plan}

With all of the above understanding and the knowledge gained from our initial attempts at tuning, the only reasonable plan remaining (before attempting to code an ASM simulator) was to approach the problem iteratively, starting with the results of the second round of individual actuator tuning and making incremental changes and noting their effect on system performance. To do this, however, we needed a quantitative method of determining the effects on settling time due to small changes made to tuning variables.

\begin{table}[t!]
    \centering
    \caption{Second group tuning configuration files}
    \vspace*{3mm}
      \begin{tabular}{c l c l} 
      \toprule
      Num & Name & Gain & Notes \\ 
      \midrule
       01 & Ring\_MaxTune &  & G1F12 \\
       02 & Ring\_MaxTune\_IntR & I & 01 w IGs red \\
       03 & Ring\_MaxTune\_IntR & I & 01 w IGs red  \\
       04 & Ring\_MaxTune2 & P & 01 w PGs inc \\
       05 & Ring\_MaxTune2\_IntR & I & 04 w IGs red \\ 
       06 & Ring\_MaxTune\_Drv01 & D & 01 w DGs at 0.02 \\
       07 & Ring\_MaxTune\_Drv02 & D & 01 w DGs at 0.05 \\
       08 & Ring\_MaxTune\_VD01 & VD & 01 w VGs at 0.02  \\
       09 & Ring\_MaxTune\-VD01 & VD & 01 w DGs at 0.05  \\
    \bottomrule
    \end{tabular}
    \label{table:grouptwo}
    \vspace{7mm}
\end{table} 

\noindent \begin{table*}[t!]
    \centering
    \caption{Results of Group One and Group Two MLAT system latency trials}
    \vspace{3mm}
      \begin{tabular}{c c l c c c c c c c } 
      \toprule
      Group & File & File & Modified & \multicolumn{3}{c}{\textbf{ Latency, \SI{}{\milli\second}}} & \multicolumn{3}{c}{\textbf{Latency, Frame Rate}} \\
      Number (G) & Number (F) & Name & Variables  & Mean & Median & Std Dev & Mean & Median & Std Dev \\ [0.5ex]
      \midrule
      2 & 04 &   Ring\_MaxTune2       & P    & 2.757	&	2.750	&	0.021	&	1.376	&	1.373	&	0.010	 \\
      2 & 07 &   Ring\_Max\_Drv02     & D    & 2.777	&	2.820	&	0.093	&	1.388	&	1.410	&	0.046	 \\
      2 & 05 &   Ring\_Max2\_IntR     & I    & 2.803	&	2.790	&	0.023	&	1.401	&	1.395	&	0.012	 \\
      2 & 03 &   Ring\_Max\_IntR2     & I    & 2.810	&	2.820	&	0.105	&	1.404	&	1.409	&	0.053	 \\
      2 & 09 &   Ring\_Max\_Vel02     & V,D  & 2.813	&	2.800	&	0.023	&	1.406	&	1.400	&	0.013	 \\
      2 & 02 &   Ring\_Max\_IntR      & I    & 2.840	&	2.840	&	0.040	&	1.420	&	1.418	&	0.020	 \\
      1 & 12 &   Ring\_MaxTune        & P    & 2.845	&	2.860	&	0.039	&	1.422	&	1.429	&	0.019	 \\
      2 & 08 &   Ring\_Max\_Vel01     & V,D  & 2.893	&	2.920	&	0.046	&	1.446	&	1.458	&	0.022	 \\
      2 & 06 &   Ring\_Max\_Drv01     & D    & 2.900	&	2.890	&	0.027	&	1.450	&	1.445	&	0.012	 \\
      1 & 10 &   Ring\_ModTune\_IntR  & P,I  & 2.909	&	2.916	&	0.054	&	1.453	&	1.458	&	0.029	 \\
      1 & 07 &   MaxTune              & P    & 2.955	&	2.974	&	0.072	&	1.477	&	1.486	&	0.036	 \\
      1 & 05 &   ModTune\_IntR        & P,I  & 2.966	&	2.954	&	0.045	&	1.482	&	1.476	&	0.023	 \\
      2 & 01 &   Ring\_MaxTune        & P    & 2.970	&	2.970	&	0.040	&	1.484	&	1.484	&	0.018	 \\
      1 & 09 &   Ring\_ModTune        & P    & 2.997	&	2.992	&	0.031	&	1.498	&	1.495	&	0.015	 \\
      1 & 01 &   CleanedCurrent       & -    & 3.082	&	3.130	&	0.134	&	1.540	&	1.564	&	0.067	 \\
      1 & 11 &   Ring\_ModTune\_IntI  & P,I  & 3.082	&	3.068	&	0.043	&	1.540	&	1.533	&	0.021	 \\
      1 & 04 &   ModTune              & P    & 3.158	&	3.184	&	0.041	&	1.579	&	1.592	&	0.020	 \\
      1 & 06 &   ModTune\_IntI        & P,I  & 3.178	&	3.178	&	0.051	&	1.588	&	1.588	&	0.025	 \\
      1 & 03 &   MinTune              & P    & 3.236	&	3.220	&	0.069	&	1.617	&	1.610	&	0.034	 \\
      1 & 08 &   Ring\_MinTune        & P    & 3.275	&	3.264	&	0.025	&	1.637	&	1.632	&	0.012	 \\
      1 & 02 &   ElwoodDefault        & -    & 3.440	&	3.476	&	0.074	&	1.720	&	1.737	&	0.037	 \\
      \bottomrule
    \end{tabular}
    \label{table:results}
    \vspace{7mm}
\end{table*} 

\subsubsection{MLAT: Measure Latency}
\label{sec:MLAT}

CACAO is the Swiss army knife of adaptive optics software. ``Compute and Control for Adaptive Optics (CACAO) is an open source software package providing a flexible framework for deploying real-time adaptive optics control.'' \cite{guyon18}. It is designed to be applicable to almost any AO system utilizing any form of hardware. MAPS, in its original laboratory testing period, had been relying on temporary control software and interfaces designed by its original software engineer. Recognizing that this initial software had limited usefulness for real-time operation, MAPS began integrating CACAO as its real-time controller software from the start of its engineering runs.

One of the core routines in CACAO is the Mirror Latency (MLAT) code. MLAT measures latency by sending DM commands (in the form of a spiral checkerboard) and monitoring when the corresponding change is seen in the output from the wavefront sensor. The operator sets the number of iterations the commands are sent. The following description is from Guyon, 2018 \citep{guyon18}:

\blockquote{ The WFS change is measured by differentiating consecutive WFS frames, as shown in Figure~\ref{fig:MLAT}. Latency is measured by finding the peak of the WFS change as a function of time delay between DM command and WFS arrival time. In order to fully sample the function, DM pokes are sent at time stamps uniformly distributed within a single WFS exposure. The peak of the function corresponds to a DM motion during the frame transfer of the WFS camera. With a frame transfer CCD and an instantaneous DM motion, the function plotted in Figure~\ref{fig:MLAT} should be a triangle of half-width equal to the WFS single frame exposure time.}

It should be noted that the MLAT code does not directly measure \emph{mirror latency}, which is the time it takes the mirror's surface figure to settle to its commanded shape after the issuance of the outer loop command. Instead it measures \emph{system hardware latency}, which includes many other factors aside from mirror latency itself.

\begin{figure*}[t!]
    \centering
        \includegraphics[height=2.8in]{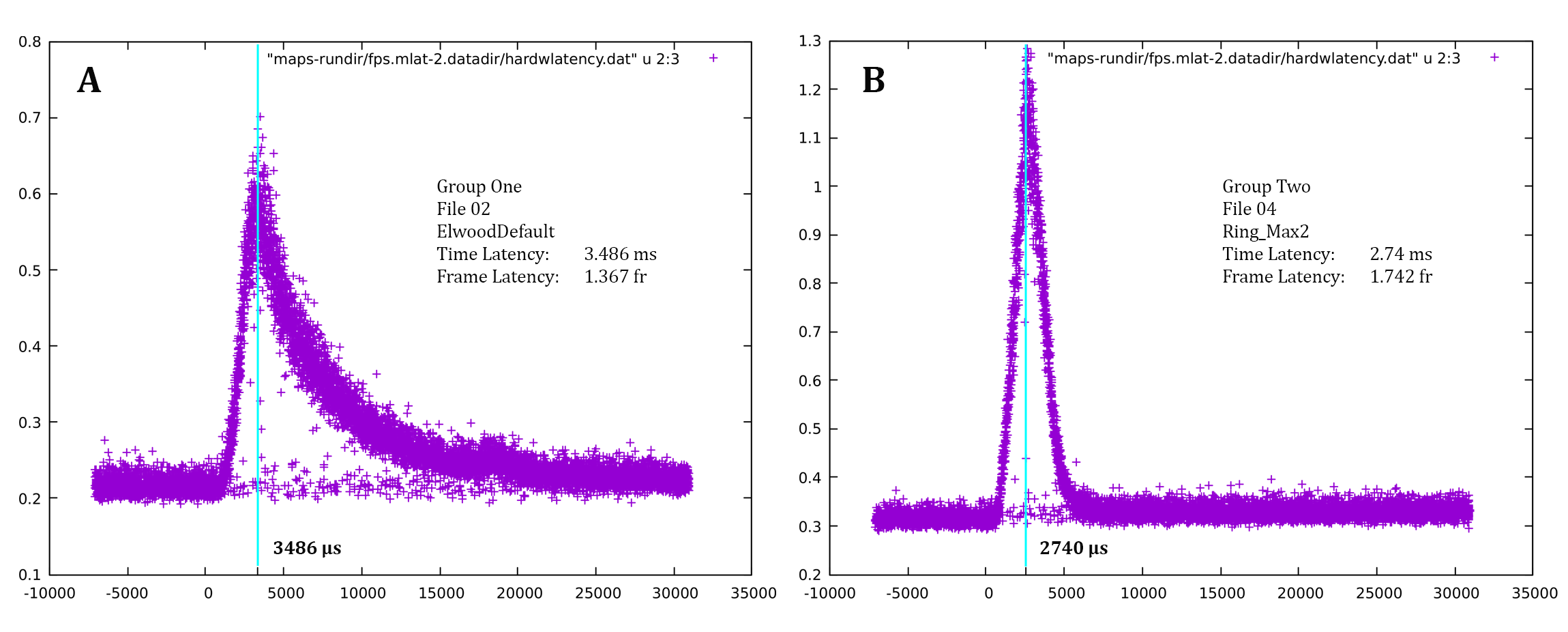}
    \caption{MLAT latency plots for the lowest and highest latency value tuning configurations. Plot A represents tuning configuration G1F02; Plot B represents tuning configuration G2F04.}
    \label{fig:latplot}
    \vspace{7mm}
\end{figure*}

But all other things being equal, MLAT is a usable metric, as mirror settling time is one of the largest contributors to overall system latency. In order to assure that this is the case, we decided tuning configurations should be tested repetitively by MLAT to determine reproducability. The smaller the variance between trials, the greater we are confident that we are seeing the result of tuning changes on mirror latency, instead of other random computational or hardware effects.

\subsubsection{Method of testing}

Tuning variables are set in MAPS in one of three ways. The first is to access the control law through the operator interface and directly enter gain values. This is configured to either access a single actuator, or access all actuators simultaneously, either of which is useful in only a narrow range of situations. The second way is through the command line, using a number of in-house commands to directly configure any combination of actuators required (say, all ring 6 actuators, numbers 90 through 125). And the third way, which is the most convenient for our purposes, is through modification of the start-up configuration files. The actuator control configuration file is a text file containing 30 settings each for 336 actuators that loads at start and determines defaults for all actuator controller settings. 

To create a tuning set, we create a configuration file with the appropriate values for the gains we wish to test. We then create multiple configuration files that incrementally change these values to determine their effect on latency. A group of these, designed to determine the optimal value for either a single or multiple tuning settings, is then tested consecutively, with the full sequence repeated multiple times.

To test a configuration, we run a script that loops through these configuration files. Each loop loads a configuration file, and then loads a flat to test the file's basic functionality. If the flat executes with minimal actuator complaint (overheating, high current draw, etc.) or system complaint (WildCoil safing, actuator errors), MLAT is run. The cycle repeats with the next file, and then through the required number of repetitions. Each file in the group is tested under the same conditions of target, system settings, etc. The process is then repeated with another group of configuration files, using the values from the file in the previous group determined to have decreased latency the most as a starting point for the new group's modifications in gain values.

Telescope time is costly, and each MLAT execution takes between two and three minutes. We decided on five repetitions of each group, with groups containing around ten different configurations. Each group's script takes approximately 2.5 hours to run. Available time to this point had allowed two configuration groups to be tested. We will continue the process as telescope time allows.

\begin{figure*}[t!]
    \centering
        \includegraphics[height=1.6in]{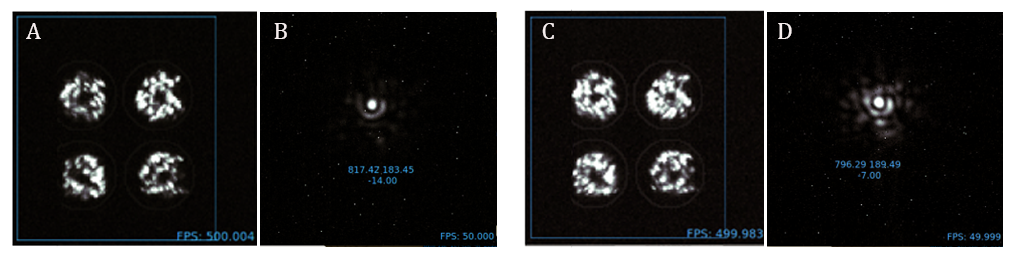}
    \caption{Images of on-sky PSFs, with their corresponding pyramid pupil images. A \& B are the results of tuning configuration G1F02, and C \& D are the results of tuning configuration G1F11.}
    \label{fig:PSF}
    \vspace{7mm}
\end{figure*}

\subsubsection{Description of first test groups}

The twenty-one initial configuration files were split into two test groups, arranged by the gain variables being tested:

\paragraph{Test Group One}

Test group one contained eleven configuration files, as given in Table~\ref{table:groupone}. These were designed to test the effects of changes to the proportional gain variable (01, 02, 03, 04, 07), the effect of changes to the integral variable (05, 06, 10, 11), and the comparison between individual actuator tuning and ring tuning (08-12). All eleven files were run consecutively, and the group was repeated five times. Date of testing was 14 August 2024. The amplitude of the command was $\SI{0.5}{\micro\metre}$ and the number of iterations was set at 500.  The target was FK5 672, Theta Herculis.

\paragraph{Test Group Two}

Test group two contained nine configuration files, all modified from Group One's Ring\_MaxTune file, as given in Table~\ref{table:grouptwo}. These were designed to test the effects of further changes to the proportional gain variable (01, 04), the effect of changes to the integral variable (02, 03, 05), the effect of changes to the derivative value (06, 07), and the effect of change to the velocity damping variable (08, 09). All nine files were run consecutively, and the group was repeated three times. Date of testing was 15 August 2024. The amplitude of the command was $\SI{0.5}{\micro\metre}$ and the number of iterations was set at 500. The target was FK5 1525, 28 Cygni.


\section{Results}

These first twenty-one test configurations were, as a whole, designed to give us a feel for the magnitude of the proportional, integral and derivative gain values, as well as to test ring tuning, and to give us an idea of both the effect of tuning on mirror latency and the reproducability of the data. 

\subsection{Latency improvement}

The data from these two sets of tests consists of a file containing the output of each DM command iteration and a plot showing the latency curve (See Section~\ref{sec:MLAT}.) The y-value of the peak point on the curve, determined by line fitting, gives the measured latency for the trial. For each set of trials for each configuration file, the mean, median and standard deviation of the latency was calculated. 

The results of all trials are shown in Table~\ref{table:results}, ordered by the latency mean from lowest to highest. The latency is expressed in both seconds and in terms of frame rate. 

All first group tests were repeated five times. The standard deviation ranges from $\sigma = 0.025$ to $\sigma = 0.134$, with a deviation mean of 0.056 and an average deviation of 0.020. All second group tests were repeated three times, with standard deviation for this group ranging from $\sigma = 0.021$ to $\sigma = 0.105$, with a deviation mean of 0.046 and an average deviation of 0.023. If there were computational or hardware issue inconsistently affecting the latency calculations, we would expect a broader variance in latency results. We take these low values as indicating that the results are replicatable and that the MLAT results are reflective of mirror latency.

These first twenty-one test configurations were, as a whole, designed to give us a feel for the magnitude of the proportional, integral and derivative gain values, as well as to test ring tuning, and to give us an idea of both the effect on mirror latency and the reproducability of the data. 

The difference between the slowest settling time, which occurred with the original Elwood Default settings (G1F02), and the fastest settling time, which occurred with large proportional gain and ring tuning (G2F04), is $\Delta t = \SI{0.683}{\milli\second}$. This represents an improvement of 20\% in the time it takes for the mirror to settle into a commanded surface figure. 

Given the approach that we are using. i.e., incrementally increasing gain values and looking for improvement, this is a very encouraging result. 

The latency plots for two of the test trials are shown in Figure~\ref{fig:latplot}. As the mirror settling time reduces, the shape of the curve more closely resembles the isoceles shape of the curve in Figure~\ref{fig:MLAT}. The broadening of the curve along the x-axis is due to timing jitter and mirror stiffness, and along the y-axis by WFS noise. 

\begin{figure*}[!t]
    \centering
        \includegraphics[height=1.6in]{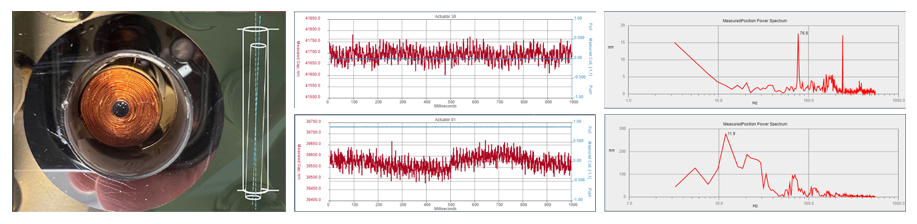}
    \caption{The effect of actuator misalignment on power spectra. Left: a top down image of a misaligned actuator in the reference body. Center: representative positional plots. Right: representative power spectra.}
    \label{fig:borehole}
    \vspace{7mm}
\end{figure*}

\subsection{Improvement in mirror modes}

The reduced system latency, assumed to be reflective of a more rapid settling time of the mirror, could be expected to manifest in improvement in the ability of the mirror to produce higher order mirror modes, as discussed in Section~\ref{sec:settling}. 

The standard operation configuration file (G1F01) allowed a maximum of 50 mirror modes. When one of the configuration files (G01F11) was used for closing the loop on sky, the mirror was able to produce up to 100 modes. 

This is to be expected, because in order to produce higher modes with higher spatial frequency corrections, the mirror must be able to push proximal segments to differing positions with large delta displacement values between them, which requires rapid settling against large mirror function damping force. This is not the case with lower order corrections, however, when proximal segments have minimal delta displacement values. 

We take this as further evidence that tuning is indeed affecting mirror settling time. 

\subsubsection{Improvement in Point Spread Function}

Although these data were received to late to analyse quantitatively, the result is visible visually. Figure~\ref{fig:PSF} shows two PSF images, along with their accompanying pyramid pupil images, from on-sky observing. The first set of images shows PSF from the G1F02 configuration. The second set of images show the PSF from the G1F11 configuration. The presence of a more complete Airy structure is visible.

\begin{figure*}[t!]
    \centering
        \includegraphics[height=1.15in]{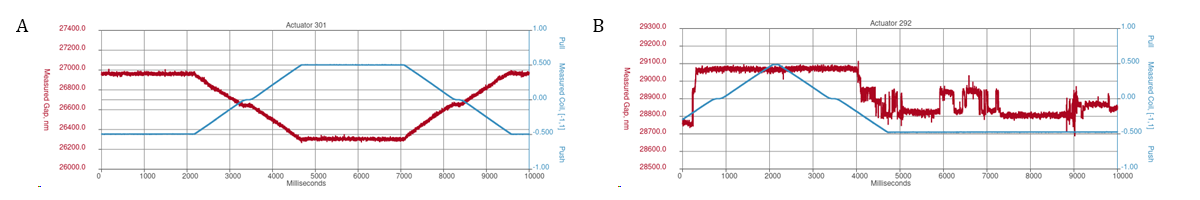}
    \caption{Position vs time plots of an actuator with suspected magnet detachment. A is a positional plot of an actuator with its corresponding mirror magnet attached responding to checkerboard motion. B is the positional plot of an actuator with a suspected detached mirror magnet.}
    \label{fig:magnet}
    \vspace{7mm}
\end{figure*}

\subsubsection{Contribution of Gain Variables}

Another result from this first round of tests is insight into the contributions that adjusting the different gain values makes towards improving latency. 

\vspace{1mm}
\noindent\textbf{Proportional Gain}: In all cases, increasing the value of the proportional gain $K_P$ decreased the latency. A maximum value has not been determined yet, nor has the interaction of the variable with other gain variables. 

\vspace{1mm}
\noindent\textbf{Integral Gain}: In all cases, decreasing the value of the integral gain $K_I$ decreased the latency. A minimum value has not been determined yet, but it has been determined that the default value is not optimal. The interaction of the variable with other gain variables, particularly the derivative gain, has not been determined.

\vspace{1mm}
\noindent\textbf{Derivative Gain}: In all cases, increasing the value of the proportional gain $K_D$ decreased the latency. A maximum value has not been determined yet, but the safe starting magnitude has been. The interaction of the variable with other gain variables, particularly the integral gain, has not been determined.

\vspace{1mm}
\noindent\textbf{Velocity Damping Gain}: In all cases, increasing the value of the velocity damping gain $K_V$ decreased the latency. A maximum value has not been determined yet, but a safe starting magnitude has been. The interaction of the variable with other gain variables, and the variance of its effect with ring position, have not been determined.

\vspace{1mm}
\noindent\textbf{Ring Tuning}: Ring tuning configurations have been determined to decrease settling time in comparison to both the individual actuator tuning configurations they were created from as well as all individual actuator tuning configurations in general.

\vspace{3mm}


\section{POSITION PLOTS AND POWER SPECTRA IN FAULT DIAGNOSIS}

\vspace{3mm}
Concurrent with our work in understanding the tuning of the MAPS controllers, we have been investigating the use of FFT power spectra in diagnosing actuator fault states. Although this work is in its beginning stages, a few useful results have already been found.

When a MAPS actuator exhibits a fault state, we might expect that there would be changes in either its positional plot or FFT power spectra, and we have tentatively found a few of these differences.

\subsection{Actuator misalignment}

When an actuator is axially misaligned with its corresponding borehole, its magnetic force no longer is normal to the mirror surface, and pulls on the segment's magnet in ways that vary with the displacement of the mirror. The resulting positional plot exhibits a low frequency oscillation, and the lower end of the power spectrum exhibits a 'tail-whipping behavior. Figure~\ref{fig:borehole} shows this behavior.

The left side of the figure is an image of an actuator and a graphic showing an axially-misaligned actuator. The right side of the figure shows both power spectra and positional plots for the actuator under conditions of maintaining a flat.  When in actuator is misaligned in this way, its positional plot will alternate between a normal looking plot (upper) and a plot that exhibits long period oscillation (lower). Its power spectrum will show its left tail whipping up and down.

Recognizing actuators that exhibit this behavior in the plots mark them for inspection during actuator maintenance, and for realignment and positioning. This correlation between alignment and plot behavior was confirmed during the rebuild discussed in Section~\ref{sec:rebuild}.

\subsection{Detached magnet}

The adhesive that holds the magnets to the thin shell has a limited lifetime, and like other ASMs that have affixed magnets (LBT, for example), magnets begin to detach. This has been a slow process, with a few magnets failing after periods of heavy ASM use. The results of magnet failure are not as obvious as one might suspect, and so determining whether a magnet has detached by inspection of an actuator's plots would be ideal.

We believe we have found indicators of magnet detachment in an actuator's positional plot. Figure~\ref{fig:magnet} shows an example. In the left side of the figure is a positional plot showing the response of a healthy actuator when it checkerboards the mirror segment. The blue current line and the red position line are inverses of each other. 

The right side of the Figure shows the positional plot of a mirror segment suspected of having a detached magnet. Although this correlation must be proven by removing the thin shell and visually inspecting the actuator, there is little else that we know of that could cause this type of behavior in the positional plot.


\section{CONCLUSION}

Starting with essentially an unknown actuator control system, we have begun to understand the magnitude ranges of tuning variables, and the effect of those variables on the overall performance of the MAPS ASM. 

We have devised an iterative system of testing configuration files which will eventually lead to optimizing mirror settling time. We will continue on-sky testing, adding additional tuning variables as telescope time allows. 

We have begun to recognize the predictive ability of power spectra analysis to diagnose actuator failure.

These efforts have already yielded a substantial increase in overall system performance, which has brought the MAPS project closer to realizing its design specifications.


\section{Acknowledgements}

The authors would like to acknowledge the substantial contributions to the MAPS adaptive secondary mirror made by control engineer Keith Powell and software engineer Elwood Downey.

The MAPS project is primarily funded through the NSF Mid-Scale Innovations Program, programs AST-1636647 and AST-1836008. 

This research has made use of NASA's Astrophysics Data System. 

We respectfully acknowledge the University of Arizona is on the land and territories of Indigenous peoples. Today, Arizona is home to 22 federally recognized tribes, with Tucson being home to the O’odham and the Yaqui. Committed to diversity and inclusion, the University strives to build sustainable relationships with sovereign Native Nations and Indigenous communities through education offerings, partnerships, and community service.)

\newpage


\bibliography{report.bib}{}
\bibliographystyle{aasjournal}

\end{document}